\documentclass[12pt]{article}

\usepackage{amsfonts, amssymb, amsthm}

\usepackage{cite}

\newcommand{\sect}[1]{\setcounter{equation}{0}\section{#1}}
\newcommand{\subsect}[1]{\subsection{#1}}

\newtheorem{tw}{Theorem}
\newtheorem{de}{Definition}
\newtheorem{co}{Corollary}
\renewcommand{\theenumi}{{\arabic{section}.\arabic{de}.\roman{enumi}}}

\newcommand{\be}{\begin{equation}}
\newcommand{\ee}{\end{equation}}
\newcommand{\bea}{\begin{eqnarray}}
\newcommand{\eea}{\end{eqnarray}}

\begin{document}


\begin{center}
{\LARGE{\bf{Geometrical origin of the  $*$-product \\ [0.40cm] in
the Fedosov formalism}}}
\end{center}

\bigskip\bigskip

\begin{center}
M. Gadella $^1$,  M.A. del Olmo $^1$ and J. Tosiek $^2$
\end{center}

\begin{center}
$^1${\sl Departamento de F\'{\i}sica Te\'orica, Universidad de Valladolid, \\
E-47011, Valladolid, Spain.}\\
\medskip

$^2${\sl  Institute of Physics, Technical University of  \L\'{o}d\'{z},\\ 
W\'{o}lcza\'{n}ska 219, 93-005 \L\'{o}d\'{z}, Poland.}\\
\medskip

{e-mail: gadella@fta.uva.es,  olmo@fta.uva.es, tosiek@p.lodz.pl}
\end{center}

\vskip 1.5cm
\centerline{\today}
\vskip 1.5cm

\begin{abstract}
The construction of the $*$-product proposed by Fedosov is implemented
in terms of the theory of fibre bundles. The geometrical origin of the
Weyl algebra and the Weyl bundle is shown. Several properties of
the  product in the Weyl algebra are proved. Symplectic and
abelian connections in the Weyl algebra bundle are introduced.
Relations between them and the symplectic connection on a phase
space $M$ are established. Elements of differential symplectic
geometry are included. Examples of the Fedosov formalism in
quantum mechanics are given.
\end{abstract}


\sect{Introduction}

The standard formulation of quantum mechanics in terms of a
complex Hilbert space and linear operators is mainly applied for
systems, whose classical limit may be described on a phase space
$\mathbb{R}^{2n}. $ There are two reasons for this situation:
 the fundamental operators $\hat{q}^i, \hat{p}_i, \; i=1, \ldots , 2n$, 
are well defined by the Dirac quantization scheme~\cite{di} only in case
$\mathbb{R}^{2n}$, and the operator orderings are based  on  the
Fourier transform~\cite{w75,co6,caj}.

Unless we agree to weaken the foregoing assumptions \cite{und} to
deal with other quantum systems it is necessary to use geometric
quantization \cite{van,woo} or deformation quantization.

Classical mechanics is a physical theory which works perfectly on
arbitrary differential manifolds. From this reason shortly after
presenting a standard version of quantum theory, researchers began
to look for an equivalent formulation of quantum mechanics based
on differential geometry.
 The first complete version of quantum theory in the language of
 the
theory of manifolds appeared in the middle of the XXth century,
when Moyal~\cite{MO49} using previous
 work by Weyl~\cite{WY31}, Wigner~\cite{WI32} and Groenewold~\cite{GW46}
presented quantum mechanics as a statistical theory. His results
are only valid for the case $\mathbb{R}^{2n}$. However, the paper
by Moyal contains the seminal ideas about deformation
quantization, since the main result of this work is the
substitution of the point-wise product of functions in phase space
for a new product which is a formal power series in $\hbar$ that
we will precise later.

A modern version of Moyal's deformation quantization on an
arbitrary differential manifold was proposed by Bayen {\it et
al}~\cite{cak} in 1978.  The mathematical structure of this
formulation of the quantum theory is based, like Hamiltonian classical
mechanics, on differential geometry of symplectic spaces.
Observables are smooth real functions on a phase space and states are
represented by   functionals. Macroworld appears in this
formalism as the limit of the quantum reality for the Planck
constant $\hbar$ tending to $0^+$.

The list of axioms constituting deformation quantization looks as
follows:
\renewcommand{\theenumi}{{\roman{enumi}}}
\begin{enumerate}
\item\label{pos1}
a state of a physical system is described on a $2n$-dimensional phase space $ M$,

\item\label{pos2}
an observable  is a real smooth function  on  $ M$,

\item\label{pos3}
for every complex-valued smooth functions $(f,g,h)$ of $C^{\infty}( M)$
the $*$-product fulfills the following conditions:
\begin{enumerate}
\item
\label{pos4}
\[
f * g= \sum_{t=0}^{\infty} \left( \frac{i \hbar}{2} \right)^t M_t(f,g),
\]
where $M_t(\cdot,\cdot)$ is a bidifferential operator on  $ M$
(see definition later, formula~(\ref{cojesz})),
 \item \label{pos5} at the classical
limit $\hbar \rightarrow 0^+$
\[
M_0(f,g)=f \cdot g;
\]
\item \label{pos6} the quasi-Dirac quantization postulate holds 
\[
M_1(f,g) - M_1(g,f)= \{f,g\}_{P},
\]
where $ \{\cdot \ , \cdot \}_{P}$ stands up for Poisson brackets;

\item \label{pos7} associativity also holds
\[
\sum_{t+u=s} (M_t(M_u(f,g),h)-M_t(f,M_u(g,h)))=0 ,\qquad \forall s \geq 0;
\]
\item \label{pos8} for the  constant function equal to $1,$ we
have
\[
M_t(1,f)=M_t(f,1)=0, \qquad \forall t \geq 1.
\]
\end{enumerate}
\end{enumerate}

Let us comment  the postulates written above.
Assumptions (\ref{pos1}) and (\ref{pos2}) say that the mathematical structure of deformation
quantization is modeled after  the classical Hamiltonian mechanics.  Differences between classical
and quantum descriptions appear at the level of  product of functions and  representation of  
states. In the quantum case the product must be  noncommutative.   The relation between this
$*$-product and the usual point-wise multiplication   of functions is established in  (\ref{pos5}).
The classical Lie algebra of functions is determined by the term $M_1$ from (\ref{pos4}).
Associativity (\ref{pos7}) is analogous to the associativity in an algebra of linear operators in
traditional quantum mechanics. Linearity of  the bidifferential operators $M_t$ plus postulate
(\ref{pos8}) expresses the fact that the measurement of a constant quantity does not interact with  
any other measurement.

Let us remember what is a bidifferential operator.  Let $M$ be a $n$-dimensional differential
manifold and   $(U, \varphi)$   a local chart on it  with 
coordinates   ${\bold q}=(q^1, \ldots, q^n)$. By a bidifferential operator  
$M_t(\cdot, \cdot) :C^{\infty}(M) \times C^{\infty}(M) \rightarrow \mathbb C $
we understand  the map which locally in a chart $(U, \varphi)$ is
a linear combination of terms like 
\be \label{cojesz}
l({\bold q}) \frac{\partial^m f({\bold q})}{(
\partial q^1)^{\alpha_1}\ldots (\partial q^n)^{\alpha_n}} \cdot
\frac{\partial^r g({\bold q})}{(
\partial q^1)^{\beta_1}\ldots ( \partial q^n)^{\beta_n}}, \qquad l,f,g \in
C^{\infty}(M), \ee 
where $\alpha_1 + \cdots + \alpha_n=m,  \;\; \beta_1 + \cdots + \beta_n=r$.  
The subindex `t' in
$M_t(\cdot, \cdot)$ denotes the highest derivative in the sum of
elements of the kind (\ref{cojesz}) in $M_t(\cdot, \cdot)$.

Note that the   assumptions of deformation quantization do not
give a construction method of   the $*$-product. In the simplest
case when the phase space is $\mathbb{R}^{2n},$ there exists a
correspondence between a wide class of linear operators acting on
vectors from a Hilbert space ${\cal H}$ of the quantum system and
functions on the phase space of it. More information about the
class of operators and functions, for which such a correspondence
exists, can be found in \cite{L2.10}. This relation, known as the
Weyl application, in the $2$-dimensional case takes the form
\cite{jap} \be \label{odw1} \hat{F} = W(F)=
\frac{1}{(2\pi)^2}\int_{\mathbb{R}^2}\alpha(\hbar\lambda \mu
)\exp\left(i[\lambda \hat{p}+\mu
\hat{q}]\right)\tilde{F}(\lambda,\mu ) d\lambda\; d\mu \ee 
where $\hat{F}$ is a linear operator acting in ${\cal H}$, $\alpha(\hbar\lambda \mu )$ is a 
function characterizing ordering of operators (however, in this work we will consider  the
case of Weyl ordering for which $\alpha(\hbar\lambda \mu )=1$), 
$p$ and $q$ are canonically conjugate coordinates on the phase space ${\mathbb R}^{2}$,  
$\hat{p}$ and $\hat{q}$ are self-adjoint operators representing momentum and position, respectively,
and fulfilling the commutation relation
\[
[\hat{q}, \hat{p}]=i \hbar 
\]
and, finally, $\tilde{F}(\lambda ,\mu )$ is the Fourier transform of the function $F(p,q)$ defined
by
\[
 \tilde{F}(\lambda ,\mu ) := \int_{\mathbb{R}^2} F(p ,q )
\exp\left(-i[\lambda p + \mu q]\right)dp\; dq.
\]
The generalization of expression (\ref{odw1}) 
to the case ${\mathbb R}^{2n}$ is straightforward.

Considering  the $*$-product as the image of the product of
operators by the inverse mapping $W^{-1}$ of the Weyl application
$W$ (with $ \alpha(\hbar\lambda \mu )=1$) we obtain that \be
\label{odw2} F *G := W^{-1}(\hat{F} \cdot \hat{G})=
 F \exp \left( \frac{i \hbar}{2}\stackrel{\leftrightarrow}{\cal P}\right) G,
\ee
where $\stackrel{\leftrightarrow}{\cal P}$ is the Poisson operator
\[
\stackrel{\leftrightarrow}{\cal P} \;:= \; \stackrel{\leftarrow }{\frac{\partial}{\partial q}}
\stackrel{\rightarrow }{\frac{\partial}{\partial p}} - \stackrel{\leftarrow }{\frac{\partial}
{\partial p}} \stackrel{\rightarrow }{\frac{\partial}{\partial q}}.
\]
The arrows  indicate the  acting direction of the partial
differential derivatives. Explicitly \be \label{odw3} F * G =
\sum_{r=0}^{\infty}\sum_{l=0}^{r}\frac{1}{r!} \left(\frac{i
\hbar}{2}\right)^r(-1)^l \frac{\partial^r F}{\partial
q^{r-l}\partial p^{l}} \frac{\partial^r G}{\partial
p^{r-l}\partial q^{l}}. \ee The mapping $W^{-1}$ is called the
Weyl correspondence and it constitutes an isomorphism between some
algebra of linear operators and an algebra of  functions on
$\mathbb{R}^{2n}$. From this fact we deduce that the $*$-product
defined as (\ref{odw2}) is associative. Expressions (\ref{odw2})
and (\ref{odw3}) are true only in
 coordinate systems in which a symplectic form $\omega$ on $\mathbb{R}^2$ is canonical.

The construction presented above cannot be used in the general case because on an arbitrary
manifold  we are not able to introduce the Fourier transform, which is fundamental for defining the
Weyl correspondence and the Weyl mapping.

As we mentioned before, Ref.~\cite{cak} does not contain an explicit
procedure for obtaining a $*$-product, only theorems of existence
of $*$-products for symplectic manifolds are presented. However, other aspects of this
paper have had   a great influence in the modern development of deformation quantization.
For instance, the deformation of the algebraic structure  of the
classical  Poisson algebra $C^{\infty}( M)$ via the $*$--product provides an example of algebra 
deformation in the sense of Gerstenhaber~\cite{Gerstenhaber}.
 Later, in the eighties, the algebraic deformation \`a la Gerstenhaber also appeared in
relation to quantum groups \cite{charipresley}. Different aspects
about  deformation quantization  and its applications  can be
found in \cite{L4}--\cite{L10}.
 For those who are
interested in general aspects of quantization deformation we
 recommend the reviews \cite{ll21,L3,L1,L1w}.

The practical realization  of deformation quantization on a
symplectic manifold was proposed by Fedosov~\cite{6,7}. His
construction is purely geometric and it is based on the theory of
fibre bundles \cite{steenrod}. Fedosov starts from some symplectic
manifold endowed  with a symmetric connection  and lifts this
connection to the so called Weyl algebra bundle. Next he builds
some new  flat connection in the Weyl bundle and shows how to
operate on flat sections of the Weyl algebra bundle in order to
define a $*$-product. In  \cite{borm,5,L15,br2,12} one can find
several considerations devoted to some aspects of the Fedosov
formalism.

The aim of this paper is to  present the Fedosov theory as an
example of application of the theory of fibre bundles in physics.
Thus, we analyze  the construction of the $*$-product, paying
attention to the geometrical aspects using the formalism of fibre
bundles. All the final results presented here are  known but we
have obtained them applying consequently  the fibre bundle
methods and in many cases  the proofs are new and   easier than
the original ones. This fact adds a pedagogical value to this work
which can complement the monographies devoted to the Fedosov theory.
As we mention before, we  are mainly interested in the geometrical
aspects of the Fedosov construction, for that reason we omit some
long and technical proofs not connected directly with geometrical
nature of the topic under study and that, moreover, can be found in
the literature.

The paper is organized as follows. In
section~\ref{symplecticgeometry} we present some results about
symplectic geometry that we will use later. The next section is
devoted to   the main mathematical structures
involved on the Fedosov method. In section~\ref{connectionsinthebundle}, 
starting from the
construction of the Weyl algebra we obtain the Weyl bundle.  This algebra is equipped with a
symplectic connection, which is studied in detail in section~\ref{starproduct}. Making  the
symplectic connection flat and defining a one-to-one mapping between 
the formal series $\Pi(\hbar,
C^{\infty}(M))$ and flat sections of the Weyl bundle, we construct a
  noncommutative associative $*$-product on $M$, this is the
content of the section~\ref{connectionsinthebundle}. Some examples
are presented in section~\ref{examples}.

\sect{Symplectic geometry}\label{symplecticgeometry}

This part is devoted to review some aspects of the geometry of
symplectic spaces. We present a procedure for defining parallel
transport for such spaces and analyze the similarities and
differences between Riemannian  and symplectic geometry.    The
reader interested in the mathematical aspects of symplectic
geometry can find a systematic presentation of this topic in
\cite{n1}. Physical applications of the symplectic geometry are
analyzed in \cite{n2,n3,caf,9}.

\begin{de}
A {symplectic manifold} $(M, \omega)$ is a manifold $M$ equipped
with a nondegenerate closed $2$-form $\omega$. The form $\omega$
is called a {symplectic form}.
\end{de}
It can be  proved (see, for instance, \cite{n1})
that the dimension of the manifold $(M, \omega)$ is always even. Hereafter, we denote it by
$\dim M = 2n$.

Let $(U_{i},\varphi_{i})$ be some local chart on $(M, \omega)$. In
this chart the symplectic form may be written as 
\be \label{now1}
\omega = \omega_{ij}dq^i \wedge dq^j, \qquad i,j = 1, 2, \ldots ,
2n. 
\ee

\begin{tw} (Darboux theorem)
\label{nowD6}
Let $(M, \omega)$ be a symplectic manifold. In the neighborhood of each point
${\tt p} \in M $ there exist local coordinates $(q^1, \ldots, q^{2n})$, 
called canonical or Darboux
coordinates, such that the form $\omega$ may be written by means of these coordinates as
\be
\label{now1.1}
\omega = dq^{n+1} \wedge dq^{1} +dq^{n+2} \wedge dq^{2} + \ldots + dq^{2n} \wedge dq^{n}.
 \ee
\end{tw}
An atlas $\{(U_{\varrho},\varphi_{\varrho})\}_{\varrho \in I}$ consisting of Darboux
charts is called the Darboux atlas on $(M, \omega)$.

As the form $\omega$ is nondegenerate, it establishes an
isomorphism $I$ between tangent $T_{\tt p}M$ and cotangent $T_{\tt
p}^*M$ spaces at each point ${\tt p} \in M$. This isomorphism $I:
T_{\tt p}M \rightarrow T_{\tt p}^*M$ is defined as follows.
\begin{de}
The $1$-form $I_X \in T_{\tt p}^*M$, with     $X \in T_{\tt p}M$,
satisfies the condition \be \label{now2} I_X(Y)=
\omega(Y,X),\qquad  \forall \: {Y \in T_{\tt p}M}   . \ee
\end{de}
    In a local chart $(U_{i},\varphi_{i})$  on $(M, \omega)$ in natural bases $\{dq^1, \ldots,
dq^{2n}\}, \{\frac{\partial }{\partial q^1}, \ldots,\frac{\partial
}{\partial q^{2n}}\}  $ of $T_{\tt p}^*M$ and $T_{\tt p}M $,
respectively, we  can write that \be
 \label{now2.5}
  (I_X)_r =
\omega_{ir}X^i.
 \ee
  The inverse mapping $I^{-1}: T_{\tt p}^*M
\rightarrow T_{\tt p}M$ can be obtained by \be \label{now2.6} X^r
= \omega^{ri}(I_X)_i, \ee where $\omega^{ri}$ is a covariant
tensor for which
 the following relation holds
\be
\label{now3}
\omega^{ij}\omega_{jk}= \delta^i_k.
\ee

The Poisson bracket  of two functions, $f,g \in C^{\infty}(M)$, is defined as
\be
\label{now4}
\{f,g\}= -\omega(I^{-1}(df),I^{-1}(dg) ).
\ee
In local coordinates
\be
\label{now5}
\{f,g\}= \omega^{ij}\frac{\partial f}{\partial q^i}\frac{\partial g}{\partial q^j}.
\ee
From the closeness of the symplectic form $\omega$, i.e. $d\omega =0$, we obtain the Jacobi identity
\be
\label{now6}
   \{\{f,g\},h\}+\{\{h,f\},g\}+\{\{g,h\},f\}=0.
\ee

In fact, it is possible to define a Poisson structure on a manifold without introducing a symplectic
form. In such cases by a Poisson structure on a manifold $M$ we understand an antisymmetric bilinear
mapping $\{\cdot, \cdot\}:
 C^{\infty}(M) \times C^{\infty}(M) \rightarrow C^{\infty}(M)$
 fulfilling the Jacobi identity (\ref{now6}) and the Leibniz rule
\[
\{f,gh\}=g \{f,h\} + \{f,g\}h.
\]
Poisson manifolds, i.e. pairs $(M, \{\cdot , \cdot\})$, are natural generalizations of symplectic
manifolds \cite{n3}.

The deformation quantization programme  works on Poisson manifolds
  \cite{br2,L24,L24a,L1q}, but we are going to continue our considerations
for the case when the phase space of a system is some symplectic manifold.

There are two ways of building a phase space for a physical
system:
\begin{itemize}
\item 
Starting from the configuration space of the system ${\cal
M}.$ At every point ${\tt p} \in {\cal M} $ we can assign the
cotangent space $T_{\tt p}^* {\cal M}.$ Taking the union
$\bigcup_{{\tt p} \in {\cal M}}T_{\tt p}^*{\cal M}$ we obtain the
cotangent bundle denoted by $T^*{\cal M}.$ Such a bundle is
equipped with the natural symplectic structure (for details see
\cite{9}). We define the phase space of the system as the
cotangent bundle $T^*{\cal M}.$ This algorithm is useful for
systems without nonholonomic constraints and it is widely used in
classical mechanics.

\item 
Having a Lie group $ G$, the
symmetry group of the physical system, we construct orbits of this Lie group
in the space of $K-$representation. These orbits can be
interpreted as phase spaces for which the given Lie group is the
group of symmetries~\cite{kiri}.

\end{itemize}

We will follow the first procedure to construct a phase space. 
Let $(M, \omega)$ be a symplectic manifold.
At every point ${\tt p} \in M $ the tangent space
$T_{\tt p}M$ is assigned.  Taking the union $\bigcup_{{\tt p} \in M}T_{\tt p}M$ we obtain the
tangent bundle denoted by $TM$.

\begin{de}
A torsion-free connection $\nabla$ in the tangent bundle $TM$ is called {symplectic}
if the induced connection $\nabla$ in $T^*M \otimes T^*M$ fulfills the condition
\be
\label{sy1}
\nabla \omega=0.
\ee
\end{de}
In a local chart $(U_{i},\varphi_{i})$ on $M$ for $ {\rm dim}\; M=2n$
the requirement (\ref{sy1}) is equivalent to the system of $2 n^2(2n-1)$   independent equations
\be
\label{sy2}
 \omega_{ij;k} := \frac{\partial \omega_{ij} }{\partial q^k} -
\Gamma^l_{ik}\omega_{lj} - \Gamma^l_{jk}\omega_{il}=0, \ee where
$\Gamma^l_{ik}$ are coefficients of the connection $\nabla$. They
are symmetric in lower indices because the connection is
torsion-free.  In a Darboux chart  all the  partial derivatives
$\frac{\partial \omega_{ij} }{\partial q^k}$ vanish, so the
equation system  (\ref{sy2}) is equivalent to the following one
\be -\Gamma_{jik} +\Gamma_{ijk}=0 \ee where
\[
\Gamma_{ijk} := \omega_{li}\Gamma^l_{jk}.
\]
\begin{co}
A connection $\nabla$ on the symplectic manifold $ (M, \omega)$ is symplectic iff
in every Darboux chart the coefficients $\Gamma_{ijk}$ are  symmetric in all the
indices $\{i,j,k\}$.
\end{co}
Unlike the Riemannian geometry on a symplectic manifold $ (M, \omega)$ we can define
many symplectic connections. The following statement holds.
\begin{tw} \cite{7} \label{99}
The symplectic connection on a symplectic manifold $(M,\omega)$ is
not unique. The set of coefficients 
\be \tilde{\Gamma}_{ijk} :=
\Delta_{ijk}+\Gamma_{ijk}, \qquad 1 \leq i,j,k \leq \dim M ,
\ee 
where
$\Delta_{ijk}$ denotes the coefficients of a tensor symmetric with
respect to indices $\{i,j,k\}$ and $\Gamma_{ijk}$ the coefficients of
a symplectic connection on $(M,\omega),$ determines a symplectic
connection on $(M,\omega).$
\end{tw}
\noindent \underline{Proof}.-  Let us consider the change of
coordinates $(q^1, \ldots q^{2n})\equiv { q}\rightarrow (Q^1,
\ldots Q^{2n})\equiv { Q}$. The transformation rule for the
coefficients $\Gamma_{ijk}$ is the following one 
\be \label{now7}
\Gamma_{ijk}({ Q}) = \Gamma_{rst}({ q}) \frac{\partial q^r}{\partial
Q^i} \frac{\partial q^s}{\partial Q^j} \frac{\partial
q^t}{\partial Q^k} + \omega_{rs}({ q}) \frac{\partial q^s}{\partial
Q^i}\frac{\partial^2 q^r}{\partial Q^j \partial Q^k}. 
\ee 
For the tensor $\Delta_{ijk} $ the relation holds 
\be \label{now7.1}
\Delta_{ijk}({ Q}) = \Delta_{rst}({q})\frac{\partial q^r}{\partial Q^i}
\frac{\partial q^s}{\partial Q^j} \frac{\partial q^t}{\partial
Q^k}. \ee From (\ref{now7}) and (\ref{now7.1}) we can see that
under the change of coordinates ${ q} \rightarrow { Q} $
 the sum $\Gamma_{ijk} + \Delta_{ijk}$ transforms under the rule (\ref{now7}), so
$\tilde{\Gamma}_{ijk} $ is a connection.\

In a Darboux chart, if $\Gamma_{ijk}$ and $\Delta_{ijk}$ are
symmetric with respect to indices $\{i,j,k\},$ the same happens
also for the sum $ \tilde{\Gamma}_{ijk}= \Gamma_{ijk} +
\Delta_{ijk}. $ It means that  $\tilde{\Gamma}_{ijk}$ is not only
a connection but also a symplectic connection.

Of course, the tensor $\Delta_{ijk}$ is symmetric in any 
coordinate system.  \hfill \rule{2mm}{2mm}

Note that locally symplectic connections exist on any symplectic manifold.
The construction of a symplectic connection on the whole manifold $(M, \omega)$ can be done
using a $C^{\infty}$-partition of unity.
\begin{de} \cite{curm}
A manifold $M$ admits $C^{\infty}$-partitions of unity if, given a locally finite
open cover $\{U_i\}_{i \in I}, $ there exists a family of $C^{\infty}$-maps $\psi_i : M \rightarrow
[0,1] $ such that the set of points at which $\psi_i$
does not vanish, ${\rm supp}(\psi_i)$, is contained in $U_i$, i.e. ${\rm supp}(\psi_i)\subset
U_i$,  and $\sum_{i \in I}\psi({\tt p})=1$ for all
${\tt p} \in M$.
\end{de}
\begin{tw}\cite{curm} \label{curm1}
Every manifold $M$ admits $C^{\infty}$-partitions of unity.
\end{tw}
Since theorem \ref{curm1} holds, on an arbitrary symplectic
manifold $(M,\omega)$ we build the symplectic connection on
$(M,\omega)$ following the   algorithm:
\renewcommand{\theenumi}{\arabic{enumi}}
\begin{enumerate}
\item
We cover the manifold $(M,\omega)$ with a Darboux atlas $\{(U_i, \varphi_i)\}_{i \in I}. $
According to theorem  \ref{curm1} there exists a $C^{\infty}$-partition of unity
$\{\psi_i\}_{i \in I}$ compatible with $\{U_i\}_{i \in I}. $
\item
In each chart  $(U_i, \varphi_i)$ we define a symplectic connection by
giving coefficients $(\Gamma_{jkl})_i$ symmetric in all the indices.
\item
The coefficients of the symplectic connection $\nabla$ on $(M,\omega)$  at an arbitrary fixed point
${\tt p} \in M $ are expressed by
\[
\Gamma_{jkl}({\tt p}) :=  \sum_{i \in I}\psi_i({\tt p}) (\Gamma_{jkl})_i.
\]
\end{enumerate}
\renewcommand{\theenumi}{{\arabic{section}.\arabic{de}.\roman{enumi}}}

\begin{de}
\label{now8} Let $(M, \omega)$ be a symplectic manifold with   a
symplectic connection $\nabla$. The triad $(M, \omega,\nabla)$ is
called {Fedosov manifold}.
\end{de}

The curvature  of a symplectic connection $\nabla$ 
is characterized by a curvature tensor.
\begin{de}
\label{now9} The {curvature tensor} $R$ of a symplectic
connection $\nabla$ in the tangent bundle $TM$ is  a mapping $R:TM
\times TM \times TM \rightarrow TM $ fulfilling the relation 
\be\label{now9.1}
 R(X,Y)Z= \nabla_X \nabla_Y Z - \nabla_Y \nabla_X Z
- \nabla_{[X,Y]}Z , \qquad \forall X,Y,Z \in TM. 
\ee
\end{de}
By $[X,Y]$ we denote, as usually, the Lie bracket of the vector
fields $X$ and $Y$. In the natural basis $\{ \frac{\partial
}{\partial q^1}  , \ldots, \frac{\partial }{\partial q^{2n}} \}$
of $TM$ the components of the curvature tensor are expressed as
\be \label{now9.2} R \left(\frac{\partial }{\partial
q^j},\frac{\partial }{\partial q^k} \right) \frac{\partial
}{\partial q^i}= R^m_{ijk}\frac{\partial }{\partial q^m}. \ee
 In
terms of the Christoffel symbols \be \label{now9.3} R^l_{ijk}=
\frac{\partial \Gamma^l_{ki}}{\partial q^j} - \frac{\partial
\Gamma^l_{ij}}{\partial q^k } + \Gamma^m_{ki} \Gamma^l_{jm} -
\Gamma^m_{ij}\Gamma^l_{km}. \ee Note that for various symplectic
connections we may have different curvature tensors. For instance,
the same symplectic manifold $(M,\omega)$  can be equipped with
flat (all the components $R^l_{ijk}$ of curvature tensor vanish) or
nonvanishing curvature.
\begin{tw}
Let us consider two Fedosov manifolds $(M, \omega, \nabla)$ and $(M,\omega, \tilde{\nabla})$
for which the relation holds $\tilde{\nabla}=\nabla + \Delta $ (see theorem \ref{99}). The
curvatures
$R $ and  $\tilde{R}$ fulfill the equation
\be
\label{now9.30}
\tilde{R}=R + r(\Delta),
\ee
where $r(\Delta)$ is some tensor depending only on $\Delta$.
\end{tw}
\noindent \underline{Proof}.-
 Since locally in a chart $(U_i,
\varphi_i)$  the equality $\tilde{\Gamma}^i_{jk}= \Gamma^i_{jk} +
\omega^{is}\Delta_{sjk} $ is fulfilled, from the relation
(\ref{now9.3}) we obtain that
\[\begin{array}{lll}
\tilde{R}^l_{ijk}&= &\frac{\partial }{\partial q^j} \left(
\Gamma^l_{ki} + \omega^{ls}\Delta_{ski}\right) - \frac{\partial
}{\partial q^k } \left( \Gamma^l_{ij}
+ \omega^{ls}\Delta_{sij}\right) \\[0.25cm]
&& \;\; + (\Gamma^m_{ki} + \omega^{ms}\Delta_{ski})
(\Gamma^l_{jm} + \omega^{ls}\Delta_{sjm}) - (\Gamma^m_{ij}
+ \omega^{ms}\Delta_{sij})(\Gamma^l_{km}+\omega^{ls}\Delta_{skm}).
\end{array}\]
It is always possible to choose the chart $(U_i, \varphi_i)$ in such a way that
at an arbitrary fixed point ${\tt p} \in M$ all  the Christoffel symbols $\Gamma^l_{jm}$ vanish.
It means that at the point ${\tt p}$ we have
\[
\tilde{R}^l_{ijk}= \frac{\partial \Gamma^l_{ki}}{\partial q^j}-
\frac{\partial \Gamma^l_{ij}}{\partial q^k } + \frac{\partial
(\omega^{ls}\Delta_{ski})}{\partial q^j}- \frac{\partial
(\omega^{ls}\Delta_{sij})}{\partial q^k } +
\omega^{ms}\Delta_{ski} \omega^{lz}\Delta_{zjm}-
\omega^{ms}\Delta_{sij}\omega^{lz}\Delta_{zkm}.
\]
So, we can write that \be \label{now9.31} \tilde{R}^l_{ijk} =
R^l_{ijk} + r(\Delta)^l_{ijk}, \ee where \be r(\Delta)^l_{ijk}=
\frac{\partial (\omega^{ls}\Delta_{ski})}{\partial q^j}-
\frac{\partial (\omega^{ls}\Delta_{sij})}{\partial q^k } +
\omega^{ms}\omega^{lz}\Delta_{ski} \Delta_{zjm}-
\omega^{ms}\omega^{lz}\Delta_{sij}\Delta_{zkm}. \ee
 \hfill \rule{2mm}{2mm}

\noindent Note that although $r(\Delta)^l_{ijk}$ looks like a
curvature tensor, in fact, it  is not a curvature because $\Delta $
is a tensor not a connection.

\vspace{0.5cm}
Lowering the upper index in $R^l_{ijk}$ we define a new tensor
\be
\label{now9.4}
R_{ijkl} := \omega_{mi}R^m_{jkl}.
\ee
This tensor has the following properties   \cite{va,gel}:
\medskip

 {\bf 1.-}
The curvature tensor is antisymmetric in the last two indices
\be
\label{now9.5}
R_{ijkl}= - R_{ijlk}.
\ee
Indeed, from (\ref{now9.3}) and the fact that the connection $\nabla$ is symmetric
we can  see that $ R^l_{ijk}= -R^l_{ikj}$.
Making the contraction (\ref{now9.4}) we obtain (\ref{now9.5}).
\hfill\rule{2mm}{2mm}
\medskip

{\bf 2.-} The first Bianchi identity holds \be \label{now9.6}
R_{ijkl} + R_{iklj} + R_{iljk}=0. \ee
Effectively, the identity
(\ref{now9.6}) is equivalent to the relation
\[
R^i_{jkl} + R^i_{klj} + R^i_{ljk}=0.
\]
It means that
\be
\label{now9.65}
R \left(\frac{\partial }{\partial q^k},\frac{\partial }{\partial q^l} \right)
\frac{\partial }{\partial q^j} +
R \left(\frac{\partial }{\partial q^l},\frac{\partial }{\partial q^j} \right)
 \frac{\partial }{\partial q^k}+
R \left(\frac{\partial }{\partial q^j},\frac{\partial }{\partial q^k} \right)
 \frac{\partial }{\partial q^l}= 0.
\ee 
On the other hand, the Lie brackets 
\be \label{now9.7} [\frac{\partial }{\partial
q^j},\frac{\partial }{\partial q^k}]=0 , \qquad j,k=1, \ldots , 2n.
\ee 
Moreover, the connection $\nabla$ is symmetric, so that
\be\label{now9.8} 
\nabla_{\frac{\partial }{\partial
q^k}}\frac{\partial }{\partial q^j} = \nabla_{\frac{\partial
}{\partial q^j}}\frac{\partial }{\partial q^k}, \qquad j,k=1,
\ldots , 2n. 
\ee 
From the definition of the curvature
(\ref{now9.1}), involving  (\ref{now9.7}) we rewrite  the l.h.s.
of (\ref{now9.65}) in the form \be\begin{array}{l} \label{now10}
\nabla_\frac{\partial }{\partial q^k}\nabla_\frac{\partial }
{\partial q^l}\frac{\partial }{\partial q^j} -
\nabla_\frac{\partial }{\partial q^l}\nabla_\frac{\partial }
{\partial q^k}\frac{\partial }{\partial q^j}+
\nabla_\frac{\partial }{\partial q^l}\nabla_\frac{\partial }
{\partial q^j}\frac{\partial }{\partial q^k}\\[0.35cm]
\qquad - \nabla_\frac{\partial }{\partial q^j}\nabla_\frac{\partial }
{\partial q^l}\frac{\partial
}{\partial q^k}+
\nabla_\frac{\partial }{\partial q^j}\nabla_\frac{\partial }{\partial q^k}
\frac{\partial }{\partial q^l} -
\nabla_\frac{\partial }{\partial q^k}\nabla_\frac{\partial }{\partial q^j}
\frac{\partial }{\partial q^l}.
\end{array}\ee
From (\ref{now9.8}) we can see that the foregoing expression vanishes.
\hfill\rule{2mm}{2mm}
\medskip

{\bf 3.-}  The second Bianchi identity is verified, i.e.
\be
\label{now11}
R_{mjkl;i} + R_{mjik;l}+R_{mjli;k}=0.
\ee
Indeed, this identity is equivalent to the relation
\be
\label{now12}
R^m_{jkl;i} + R^m_{jik;l}+R^m_{jli;k}=0.
\ee
It is always possible to consider locally such a chart on $(M, \omega, \nabla)$
that at an arbitrary fixed point ${\tt p} \in M$ all the coefficients $\Gamma^i_{jk}$ disappear. If
we compute all the covariant derivatives from (\ref{now12}) at ${\tt p},$ we obtain that the l.h.s.
of (\ref{now12}) equals
\[
\frac{\partial^2 \Gamma^m_{lj}}{\partial q^i \partial q^k } -
\frac{\partial^2 \Gamma^m_{kj}}{\partial q^i \partial q^l } +
\frac{\partial^2 \Gamma^m_{kj}}{\partial q^l \partial q^i } -
 \frac{\partial^2 \Gamma^m_{ji}}{\partial q^l \partial q^k } +
\frac{\partial^2 \Gamma^m_{ij}}{\partial q^k \partial q^l } -
 \frac{\partial^2 \Gamma^m_{lj}}{\partial q^k \partial q^i }=0.
\]
\hfill\rule{2mm}{2mm}

It is easy to see that in fact properties 1--3 hold for any curvature tensor of
a symmetric connection.
Now we are going to present some features of the symplectic curvature tensors exclusively \cite{va}.
\medskip

{\bf 4.-}
The symplectic curvature tensor $R_{ijkl}$ is symmetric in the first two indices, i.e.
\be
\label{now13}
R_{ijkl}= R_{jikl}.
\ee
The proof is as follows. From the Darboux theorem we can always find a system of coordinates in
which  the symplectic tensor $\omega_{ij}$ is locally constant. In such a chart we have
\be
\label{now14}
R_{ijkl}= \frac{\partial \Gamma_{ilj}}{\partial q^k} -
 \frac{\partial \Gamma_{ijk}}{\partial q^l} + \omega^{zu}\Gamma_{ulj}\Gamma_{ikz}-
\omega^{zu}\Gamma_{ujk}\Gamma_{ilz}
\ee
\be
\label{now15}
R_{jikl}= \frac{\partial \Gamma_{jli}}{\partial q^k} -
 \frac{\partial \Gamma_{jik}}{\partial q^l} + \omega^{zu}\Gamma_{uli}\Gamma_{jkz}-
\omega^{zu}\Gamma_{uik}\Gamma_{jlz}.
\ee
Permuting repeated indices $z \leftrightarrow u$ in (\ref{now15}) we see that
\[
R_{jikl}= \frac{\partial \Gamma_{jli}}{\partial q^k} -
\frac{\partial \Gamma_{jik}}{\partial q^l} + \omega^{uz}\Gamma_{zli}\Gamma_{jku}-
\omega^{uz}\Gamma_{zik}\Gamma_{jlu}.
\]
Remembering that $\omega^{zu}= - \omega^{uz}$ and using the fact
that in Darboux coordinates the coefficients $\Gamma_{jli}$ are symmetric
in all the indices we obtain that
$R_{ijkl}= R_{jikl}$.
\hfill\rule{2mm}{2mm}
\medskip

{\bf 5.-}
For a symplectic curvature tensor the following relation holds
\be
\label{now16}
R_{ijkl}+R_{jkli}+ R_{klij}+R_{lijk}=0.
\ee
Indeed, from the first Bianchi identity (\ref{now9.6}) we get
\be\begin{array}{l}
\label{p1}
R_{ijkl}+R_{iljk}+R_{iklj}=0,\\[0.25cm]
R_{jikl}+R_{jlik}+R_{jkli}=0,\\[0.25cm]
R_{kijl}+R_{kjli}+R_{klij}=0,\\[0.25cm]
 R_{lijk}+R_{lkij}+R_{ljki}=0.
\end{array}\ee
From (\ref{now9.5}) and (\ref{now13}) adding formulas (\ref{p1})
we conclude that the equation (\ref{now14}) is always true for any symplectic curvature.
\hfill\rule{2mm}{2mm}
\medskip

Let us compute the number of independent components of the tensor
$R_{ijkl}$. As usually, we assume that ${\rm dim}\; M = 2n$. From
the properties of antisymmetry (\ref{now9.5}) and symmetry
(\ref{now13}) we obtain that the tensor $R_{ijkl}$ has
$n^2(2n+1)(2n-1)$ independent elements. The first Bianchi identity
(\ref{now9.6}) is a new constraint if and only if all the  indices
$j,k,l$ are different in $R_{ijkl}$. It means that (\ref{now9.6})
provides us $2n \left( \begin{array}{c} 2n \\ 3 \end{array}
\right)$ new equations. The property (\ref{now16}) can no be reduced by
the symmetry of the tensor or by the first Bianchi identity only if
all the indices  $i,j,k,l$ are different, so we have $\left(
\begin{array}{c} 2n \\ 4
\end{array} \right)$
 independent conditions. Hence, the
symplectic curvature tensor $R_{ijkl}$  has
 \[
n(2n-1)\left( n(2n+1)- \frac{2}{3}n(2n-2)- \frac{1}{12}(2n-2)(2n-3)\right)
\]
independent components.
For example, on a $2$-dimensional Fedosov manifold $(M, \omega, \nabla)$
its curvature tensor has 3 independent components. For ${\rm dim}\; M=4$ the amount of
independent elements of $R_{ijkl}$ increases to $47$. When $M $ is a Riemannian manifold,
the number of independent components of a curvature tensor  would be equal to $1$ and $20$,
respectively.

Apart from the curvature tensor $R_{ijkl}$ the geometry of a symplectic space  can
be characterized by a Ricci tensor
and a scalar of curvature.
 \begin{de}
The Ricci tensor on a Fedosov manifold $(M,\omega, \nabla)$ is
defined by 
\be \label{now20} 
K_{ij}:= \omega^{kl}R_{likj}= R^k_{ikj}. 
\ee
\end{de}
The Ricci tensor is symmetric, i.e.
\be
\label{now21}
K_{ij}=K_{ji}.
\ee
Indeed, from (\ref{now9.5}), (\ref{now13}) and (\ref{now16}) we can write
\[
\omega^{ki}(R_{ijkl}+R_{jkli}+ R_{klij}+R_{lijk})= K_{jl} +  K_{jl} - K_{lj} - K_{lj}=0 .
\]
So, the equality (\ref{now21}) holds.
\hfill\rule{2mm}{2mm}
\begin{co}
On any Fedosov manifold
\be
\label{now22}
\omega^{ji}K_{ij}=0.
\ee
\end{co}
This conclusion is a straightforward consequence of the fact that
the Ricci tensor $K_{ij}$ is symmetric and $\omega^{ji}$
antisymmetric. From (\ref{now22}) we can see that the scalar of
any symplectic curvature defined  as $K:=\omega^{ji}K_{ij}$ is
always $0$.


\sect{The Weyl bundle}\label{Weylbundle}

In this section we present the construction of a bundle, which plays a fundamental role in
the definition of the $*$-product on a symplectic manifold. Let us start defining a
formal series over a vector space.
\begin{de} \cite{5}
Let $\lambda$ be a fixed real number and $ V $ some vector space.
A formal series  in the formal parameter $\lambda$ is  each
expression of the form \be \label{1} 
v[[\lambda]]=
\sum_{i=0}^{\infty}\lambda^i v_i, \qquad \forall v_i \in V  . \ee
\end{de}
The set of formal series $v[[\lambda]]$ constitutes a vector
space. Addition means vector summation
of the elements of the same power of $ \lambda$, and
multiplication by a scalar, $a$, is multiplication of each vector standing on the r.h.s. of
(\ref{1})  by $a$, i.e.
\[\begin{array}{l}
u[[\lambda]] + v[[\lambda]] = \sum_{i=0}^{\infty}\lambda^i (u_i + v_i) \\[0.25cm]
a \cdot v[[\lambda]]= \sum_{i=0}^{\infty}\lambda^i (a v_i).
\end{array}\]
The vector space of formal series  over the vector space $ V $ in the parameter $\lambda$ 
(\ref{1})   will be  denoted by $V[[\lambda]]$ and can be considered as a direct sum
\be
\label{11}
V[[\lambda]] = \bigoplus_{i=1}^{\infty} V_i,\qquad V_i=V\;\;\; {\forall i}.
\ee

Let $(M,\omega)$ be a symplectic manifold and $T^*_{\tt p}M$  the cotangent space to $M$
at the point ${\tt p}$ of $M$.
 The space $(T^*_{\tt p}M)^l$  is a symmetrized tensor product of
${T^*_{\tt p}M \otimes ^{\rm l - times} \ldots  \otimes T^*_{\tt p}M} $.
It is spanned by
\be
\label{now01}
\underbrace{v_{i_1} \otimes \ldots \otimes
v_{i_l}}_{\rm symmetrized} := \frac{1}{l!}
\sum_{\rm all \; permutations} v_{\sigma i_1} \otimes
\ldots \otimes
v_{\sigma i_l},
\ee
where $v_{i_1}, \ldots,v_{i_l} \in  T^*_{\tt p}M$.

\begin{de}
A {preweyl vector space} $P^*_{\tt p}M$ at the point ${\tt p} \in M$ is the direct sum
\[
P^*_{\tt p}M :=
\bigoplus_{l=0}^{\infty}(T^*_{\tt p}M)^l.
\]
\end{de}
We introduce the formal series over the preweyl vector space as follows.
\begin{de}
A {Weyl vector space} $P^*_{\tt p}M[[\hbar]]$ is the vector
space of formal series over the preweyl vector space  $P^*_{\tt
p}M$ in the formal parameter $\hbar$.
\end{de}
For further physical applications we usually  identify $\hbar$ with
the Planck constant. The elements of $P^*_{\tt p}M[[\hbar]]$  can
be written in the form
\be \label{2}
 v_{\tt p}[[\hbar]]= \sum_{k=0}^{\infty}
\sum_{l=0}^{\infty} \hbar^k a_{k,i_1 \ldots i_l} ,\qquad
a_{k,i_1 \ldots i_l} \in P^*_{\tt p}M.
\ee
For $l=0 $ we
have just the sum  $ \sum_{k=0}^{\infty} \hbar^k a_{k}$.
\begin{de} \cite{7}
The {degree} of the component $ a_{k,i_1 \ldots i_l} $, ${\rm deg}\; a_{k,i_1 \ldots i_l}  $, 
of the  Weyl vector space $ P^*_{\tt p}M[[\hbar]]$  is  $2k+l$.
\end{de}

Our aim is to define a product (denoted by  $\circ $)  of elements of the Weyl space
$ P^*_{\tt p}M[[\hbar]]$ which  equips the Weyl space with an algebra structure. Such a product
must give a symmetric tensor. Moreover, we require that ${\rm deg} (a \circ  b )=  {\rm deg}\; a+
{\rm deg}\; b$.

Let us assume that the elements of the Weyl vector space are written in a natural basis constructed
in terms of  symmetric tensor products of $ \{dq^1, \ldots ,dq^{2n} \},$ where $\dim M =2n$, and
$X_{\tt p}
\in T_{\tt p}M$ is some fixed vector of the  space $ T_{\tt p}M$    tangent to $M$ at the point
${\tt p}$. Let us denote the components of $X$ in the basis $\{\frac{\partial}{\partial q^1},
\ldots,
\frac{\partial}{\partial q^{2n}} \}$  by $ X_{\tt p}^i$.

It is obvious that for every $a_{k,i_1 \ldots i_l} \in P^*_{\tt p}M[[\hbar]]$
\[
a_{k,i_1 \ldots i_l}(\underbrace{X_{\tt p},
\ldots,X_{\tt p}}_{\rm l - times})=
a_{k,i_1 \ldots i_l}X_{\tt p}^{i_1} \cdots X_{\tt p}^{i_l}
\]
is a complex number and we can handle $a_{k,i_1 \ldots
i_l}({X_{\tt p}, \ldots,X_{\tt p}})$ as a polynomial of  $l$-th degree in the components of
the vector $X_{\tt p}$. We  extend this observation to each
element of the Weyl algebra $P^*_{\tt p}M[[\hbar]]$ and consider
$v_{\tt p}[[\hbar]](X_{\tt p} ) $ like a function of $X_{\tt p}^1,
\ldots , X_{\tt p}^{2n}$ of the form \be \label{now001} v_{\tt
p}[[\hbar]](X_{\tt p} )= \sum_{k=0}^{\infty} \sum_{l=0}^{\infty}
\hbar^k a_{k,i_1 \ldots i_l} X_{\tt p}^{i_1} \cdots X_{\tt
p}^{i_l}. \ee We do not define any topology in the space of
functions of the kind (\ref{now001}).

By the derivative $\frac{\partial v_{\tt p}[[\hbar]]}{\partial
X_{\tt p}^{i}} $ we understand the derivative of the sum (\ref{now001}) as function of $X_{\tt p}^1,
\ldots , X_{\tt p}^{2n}$ .

\begin{de}   \cite{6,7}
The product $\circ: P^*_{\tt p}M[[\hbar]] \times P^*_{\tt p}M[[\hbar]]
\rightarrow
P^*_{\tt p}M[[\hbar]]
$ of two elements $a, b \in P^*_{\tt p}M[[\hbar]]$ is an element  $ c \in P^*_{\tt p}M[[\hbar]]$
such that for each $X_{\tt p} \in T_{\tt p}M$ the following equality holds
\be
\label{6}
c(X_{\tt p})= a(X_{\tt p}) \circ b(X_{\tt p}) :=  \sum_{t=0}^{\infty} \frac{1}{t!}\left(
\frac{i
\hbar}{2}\right)^t
\omega^{i_1 j_1} \cdots \omega^{i_l j_l}
 \frac{\partial^t \; a(X_{\tt p}) }{\partial X^{i_1}_{\tt p}\ldots
\partial X^{i_l}_{\tt p} }
\frac{\partial^t \; b(X_{\tt p}) }{\partial X^{j_1}_{\tt p}\ldots
\partial X^{j_l}_{\tt p} }.
\ee
\end{de}
The pair $(P^*_{\tt p}M[[\hbar]],\circ) $ is a noncommutative associative
algebra called the {Weyl algebra} and denoted by $ {\cal P^*M}_{\tt p}[[\hbar]]$.
Let us simply enumerate some properties of the $\circ$-product:
\renewcommand{\theenumi}{\bf \arabic{enumi}}
\begin{enumerate}
\item
The $\circ$-product is independent of the chart. 
This result is a straightforward consequence
of the fact that all
the elements appearing on the r.h.s. of the formula (\ref{6}) are scalars.
\item
 The $\circ$-product is associative but in general nonabelian.
The associativity of the  $\circ$-product may be roughly explained
in the following way. The elements $a(X_{\tt p}) $ and $b(X_{\tt
p})$ are polynomials in $X^i$'s. Let us formally substitute $X^i
\rightarrow q^i,$ where by $q^i$ we denote the cartesian
coordinates on the symplectic space $(\mathbb{R}^{2n},\omega)$.
After this substitution the product    for polynomials in
variables $q^i$'s defined like (\ref{6}) is a associative Moyal
product (see the relation (\ref{odw2})). Indeed, in Darboux
coordinates in the $2$-dimensional case with the symplectic form
$\omega= dq^2 \wedge dq^1$, the formula (\ref{6}) may be written
as
\[
a \circ b = \sum_{t=0}^{\infty} \frac{1}{t!}\left(
\frac{i
\hbar}{2}\right)^t \sum_{l=0}^{t}(-1)^l
\frac{\partial^t \; a(X_{\tt p}) }{\partial (X^{1}_{\tt p})^{t-l}
\partial (X^{2}_{\tt p})^{l} }\;
\frac{\partial^t \; b(X_{\tt p}) }{\partial (X^{1}_{\tt p})^{l}
\partial (X^{2}_{\tt p})^{t-l} }
\]
which is exactly (\ref{odw3}). 
\item \label{znowu} 
The degree verifies that 
\be\label{degree}
{\rm deg}\; (a \circ b )= {\rm deg}\; a + {\rm deg}\; b. 
\ee
 This
statement, according to the definition (\ref{6}), is obvious.
\end{enumerate}
\vspace{0.25cm}

In order to enlighten the above construction of the 
$\circ$-product let us consider the following
example.
Let $(M, \omega)$ be  a $2$-dimensional symplectic manifold.
From the Darboux theorem (\ref{nowD6}) we can choose a
chart $(U_i,
\varphi_i)$ in such a manner that at a fixed point ${\tt p} \in M$ and in
some neighborhood of it we have that
\[
\omega= dq^2 \wedge dq^1.
\]
It means that a chart $(U_i, \varphi_i)$  in a natural basis
$\{\frac{\partial }{\partial q^1},\frac{\partial }{\partial q^2}
\}$ the tensor $\omega^{ij}$ takes the form
\[
\omega^{ij}=
\left(
\begin{array}{cc}
0 & 1 \\
-1 & 0
\end{array}
\right).
\]
Let us consider the $\circ$-product of elements $a= a_{0,[11]}$ and $b=b_{0,[12]}$, where   the
symbol $[\ldots]$   denotes the symmetrization in the indices inside the bracket. From (\ref{6}) we
obtain
\[\begin{array}{lll}
a(X)\; \circ \;b(X) &=& a_{0,11}X^1X^1 \;\circ \;b_{0,12}X^1 X^2 \\[0.25cm]
&=&
a_{0,11}b_{0,12}X^1X^1X^1X^2  + \frac{1}{1!}\left( \frac{i \hbar}{2}\right) \cdot 1 \cdot 2
a_{0,11}b_{0,12} X^1X^1 .
\end{array}\]
So, finally
\[
c_{0,[1112]}=a_{0,11}b_{0,12}\qquad
 c_{1,[11] }=i \cdot a_{0,11}b_{0,12}.
\]

\vspace{0.5cm}
We have worked until now  at an arbitrary fixed point ${\tt p}$  of
the symplectic manifold $(M,\omega)$.
 But our aim is to define the $*$-product on the whole manifold. To achieve it we need
to introduce a new object, the so called Weyl algebra bundle.

Let us start reminding  the definition of a fibre bundle.
\begin{de} \cite{10}
A (differentiable) {fibre bundle} $E\equiv (E, \pi,M,F,G)$ consists of
the following elements:
(1) the  differentiable manifolds $E, M$ and $F$ called the {   total space},
 the {  base space} and the { fibre} (or {  typical
fibre}), respectively; (2) 
a  surjection $\pi:E \rightarrow M$ called the {  projection}. The
inverse image $\pi^{-1}({\tt p}) \equiv F_{\tt p}\cong F$ is called the
fibre at ${\tt p}\in M$;
(3) a Lie group $G$ called the {  structure group}, which acts on $F$ at
the left; (4) a set of open coverings $\{U_i\}$ of $M$ with a diffeomorphism
$\varphi_i:U_i \times F \rightarrow \pi^{-1}(U_i) $ such that
$\pi\varphi_i ({\tt p},f)={\tt p}. $ The map $\varphi_i$ is called
the {local trivialization} since $\varphi_i^{-1}$ maps $\pi^{-1}(U_i)$
onto the direct product $U_i \times F$;
(5) if we write $\varphi_i ({\tt p},f)=\varphi_{i,{\tt p}} (f),$ the map
$\varphi_{i,{\tt p}}:F \rightarrow F_{\tt p}$ is a diffeomorphism. On
$U_i \cap U_j \neq \emptyset,$ we require that $t_{ij}({\tt p}) := 
\varphi_{i,{\tt p}}^{-1}\varphi_{j,{\tt p}}:F \rightarrow F$ be an
element of $G$. Then $\varphi_{i}$ and $\varphi_{j}$ are related by a
smooth map $t_{ij}:U_i \cap U_j \rightarrow G$ as
\[
\varphi_j ({\tt p},f)=\varphi_i ({\tt p},t_{ij}({\tt p})f).
\]
The elements $\{t_{ij}\}$ are called the { transition functions}.
\end{de}

A section of the fibre bundle is a map $s : M\to E$ such that $(\Pi \circ  s )({\tt p}) ={\tt p}$ for
every $\tt p$ of $M$. The   space of all the (smooth) sections will be denoted by $C^{\infty}(M,E)$
or simply
$C^{\infty}(E)$.

A differential fibre bundle where the typical fibre $F$ and the   fibres $F_{\tt p}$ are vector
spaces  is called a vector bundle. Obviously, if $F$ and the   fibres $F_{\tt p}$ have the structure
of algebra we have an algebra bundle.

 \begin{de}
\label{wiazkaWeyla}
The collection of all the Weyl algebras, i.e.
\be
{\cal P^*M}{}[[\hbar]] = \bigcup_{{\tt p} \in M} {\cal P^*M}_{\tt p}[[\hbar]] ,
\ee
is a vector bundle which is also an algebra bundle and it is called the {Weyl algebra bundle}.
\end{de}

The structure of the Weyl algebra bundle looks as follows:
\begin{itemize}
\item
The set ${\cal P^*M}{}[[\hbar]]$ is the total space.
\item
The symplectic space $ (M,\omega) $ is the base space.
\item
The fibre is the Weyl algebra ${\cal P^*M}_{\tt p}[[\hbar]]$. We do not introduce a new symbol but
now ${\cal P^*M}_{\tt p}[[\hbar]]$ is not connected with any point. Moreover, ${\cal P^*M}_{\tt
p}[[\hbar]]$ is a Weyl vector space  and
$ \omega^{ij},\;\; i,j= 1, \ldots , 2n$,
are the coefficients of a fixed nondegenerate antisymmetric tensor.
\item
The projection $\pi:{\cal P^*M}{}[[\hbar]] \rightarrow M $ is defined by
$\Pi (v_{\tt p}[[\hbar]])={\tt p}$.
\item
 $GL(2n,\mathbb{R})$ is the group of real automorphisms of  $T^*_{\tt p}M$.
The structure group of the fibre is 
\be \label{sg1} 
G := \bigoplus_{l=0}^{\infty}(\underbrace{GL(2n,\mathbb{R}) \otimes
\ldots \otimes GL(2n,\mathbb{R})}_{\rm l-times }). 
\ee 
For $l=0$
we have just the identity transformation. The tensor $\omega^{ij}$
transforms under the group $GL(2n,\mathbb{R}) \otimes
GL(2n,\mathbb{R}) $. Moreover, if the tensor $a_{ij}$   transforms
under the element $g \in GL(2n,\mathbb{R}) \otimes
GL(2n,\mathbb{R}),$ then also $\omega^{ij}$ transforms under
$g^{-1} \in GL(2n,\mathbb{R}) \otimes GL(2n,\mathbb{R})$. \item
Let $(U_i,\phi_i)$ be a chart on $M$. The local trivialization
$\varphi_i$ is a diffeomorphism which assigns to every point of
${\cal P^*M}[[\hbar]]$  the point $ {\tt p}$ of $M$
 and to the element of ${\cal P^*M}_{\tt p}[[\hbar]]$ its coordinates in the natural basis
$\{\frac{\partial}{\partial q_{i_1}} \otimes \cdots \otimes \frac{\partial}
{\partial q_{i_l}}\}$ determined by
$\phi_i$.
\end{itemize}

Other kind of  fibre bundle used in this work is the
cotangent bundle.

\begin{de} \cite{10}
The {\bf cotangent bundle} $T^*M$ of a differentiable manifold $M$ is
\be
T^*M \equiv \bigcup_{{\tt p} \in M} T^*_{\tt p}M,
\ee
where $T^*_{\tt p}M$ is the cotangent space at the point  ${\tt p}$ of $M$.
\end{de}

Let $\Lambda^k$ be  the vector bundle
\[
(\underbrace{T^*M \otimes \ldots \otimes T^*M}_{{\rm k-times} \; +
\; {\rm
antisymmetrization}},\pi,M,T^*M,\underbrace{GL(2n,\mathbb{R})
\otimes \ldots \otimes GL(2n,\mathbb{R})}_{\rm k- times} )
\]
of  $k$-forms on $M$. Taking the direct sum of tensor products of bundles
\[
{\cal P^*M}[[\hbar]] \Lambda = \bigoplus_{k=0}^{2n} {\cal P^*M}[[\hbar]] \otimes \Lambda^k
\]
we obtain a new algebra bundle.
A product  can be defined in it in terms of the  $\circ$-product of elements of
the Weyl algebra
${\cal P^*M}_{\tt p}[[\hbar]]$ and the external product  ($\wedge$)  of forms. We will denote the
new product also by `$\circ$'. In a local chart $(U_i, \varphi_i)$ the elements of ${\cal
P^*M}[[\hbar]]
\otimes
\Lambda^k$ are smooth tensor fields of the kind
\[
a_{c,i_1 \ldots i_l, j_1 \ldots j_k}(q^1, \ldots, q^{2n}).
\]
These objects $a_{c,i_1 \ldots i_l, j_1 \ldots j_k}$ are symmetric in the indices $(i_1 \ldots i_l)$
(as elements of the Weyl algebra) and antisymmetric in $(j_1, \ldots, j_k)$ (as forms). For
simplicity we will omit coordinates $(q^1, \ldots, q^{2n})$.  The elements of ${\cal P^*M}[[\hbar]]
\Lambda$   can be seen as forms with values in the Weyl algebra.

In the special case when $a \in \Lambda^{0}$ is a smooth function on $ M$, we obtain
\[
a \circ b = a \cdot b = a \wedge b , \qquad  \forall\; b \in {\cal P^*M}[[\hbar]]  \Lambda.
\]

\begin{de} \cite{12}
\label{def1}
The { commutator} of two forms $a \in {\cal P^*M}[[\hbar]] \otimes \Lambda^{k_1}$ and $b \in
{\cal P^*M}[[\hbar]] \otimes \Lambda^{k_2}$ is a form belonging to
$ {\cal P^*M}[[\hbar]] \otimes \Lambda^{k_1 + k_2}$ such that
\be
\label{kom}
[a,b] :=  a \circ b -(-1)^{k_1 \cdot k_2}b \circ a.
\ee
\end{de}
We point out some properties of the commutator of forms with values in the Weyl algebra:
\renewcommand{\theenumi}{\bf \arabic{enumi}}
\begin{enumerate}
\item
The straightforward consequence of (\ref{kom}) is the equality
\be
\label{kom1}
[b,a]=(-1)^{k_1 \cdot k_2 + 1}[a,b].
\ee
\item
The Jacobi identity
\be
\label{kom01}
(-1)^{k_2(k_1 + k_3)}[a,[b,c]]+(-1)^{k_1(k_2 + k_3)}[c,[a,b]]+(-1)^{k_3(k_1 + k_2)}[b,[c,a]]=0
\ee
holds for   every  $a \in  {\cal P^*M}[[\hbar]] \otimes \Lambda^{k_1}$, $b \in  {\cal
P^*M}[[\hbar]] \otimes \Lambda^{k_2}$ and  $c \in  {\cal P^*M}[[\hbar]]
\otimes \Lambda^{k_3}$. It is a simple  matter of computation to prove this identity.

\item
\be
\label{komu1}
[a,b \circ c] =(-1)^{k_1 k_2} b \circ [a,c] + [a,b]\circ c
\ee
for all $a \in  {\cal P^*M}[[\hbar]] \otimes \Lambda^{k_1}$, $b \in  {\cal
P^*M}[[\hbar]] \otimes \Lambda^{k_2}$ and $ c \in {\cal P^*M}[[\hbar]] \otimes
\Lambda^{k_3}$.

\underline{Proof}.-
 Indeed, \be \label{komu2} [a,b \circ c] = a
\circ b \circ c -(-1)^{k_1(k_2 + k_3)} b \circ c \circ a. 
\ee 
Developing   the r.h.s. of (\ref{komu1})  we get 
\be \label{komu3} (-1)^{k_1 k_2} b \circ
[a,c]=(-1)^{k_1 k_2}b \circ a \circ c - (-1)^{k_1 k_2}(-1)^{k_1
k_3} b \circ c \circ a 
\ee 
and 
\be \label{komu4} [a,b]\circ c = a
\circ b \circ c - (-1)^{k_1 k_2} b \circ a \circ c. 
\ee 
Summing
(\ref{komu3}) and (\ref{komu4}) we recover (\ref{komu2}).
\hfill\rule{2mm}{2mm}

\item
\label{co1}
The commutator   $[a,b]$ of two forms $a \in  {\cal P^*M}[[\hbar]] \otimes \Lambda^{m} $
and $b \in  {\cal P^*M}[[\hbar]] \otimes \Lambda^{n}$ contains only 
terms with odd number of derivatives in $X^i$.
\item
\label{co2}
The commutator $[a,b]$ of two real forms $a \in  {\cal P^*M}[[\hbar]] \otimes \Lambda^{m} $
and $b \in  {\cal P^*M}[[\hbar]] \otimes \Lambda^{n}$ is purely imaginary.

Proofs of these last two  properties  can be found in \cite{my}.
\end{enumerate}


\sect{Connections in  the bundle ${\cal P^*M}[[\hbar]]
\Lambda$}\label{connectionsinthebundle}

Let $(M,\omega, \nabla)$ be  a Fedosov manifold    equipped
with a symplectic connection $\nabla$, whose coefficients in the Darboux
atlas $\{(U_{\varrho},\varphi_{\varrho})\}_{\varrho \in I}$ are
$\Gamma_{ijk}$. The connection allows us to know how to transport
geometrical quantities parallelly on $M$.

Let us define the  covariant derivative of a tensor 
$a_{t,i_1 \ldots i_l,j_1 \ldots j_k}$ with
respect the variable $q^s$.
 Following  \cite{10} we can write
\[\begin{array}{l}\label{x1}
\partial_{;s} a_{t,i_1 \ldots i_l,j_1 \ldots j_k}: =
\frac{\partial a_{t,i_1 \ldots i_l,j_1 \ldots
j_k}}{\partial q^s}-
\omega^{ur}\Gamma_{rsi_1}a_{t,u i_2 \ldots i_l,j_1, \ldots j_k} - \ldots  \\[0.25cm]
\qquad -\; \omega^{ur}\Gamma_{rsi_l}a_{t,i_1  \ldots i_{l-1}u,j_1, \ldots j_k}
 - \omega^{ur}\Gamma_{rsj_1}a_{t, i_1 \ldots i_l,u j_2 \ldots j_k}- \ldots -
\omega^{ur}\Gamma_{rsj_k}a_{t, i_1 \ldots i_l,j_1 \ldots j_{k-1}u}.
\end{array}\]
The covariant derivative $\partial_{;s}$ increases the covariance
of the tensor $a_{t,i_1 \ldots i_l,j_1 \ldots j_k}$, so the object
$\partial_{;s} a_{t,i_1 \ldots i_l,j_1 \ldots j_k}$ is a tensor of
the range $(0,l+k+1)$. On the other hand, the covariant derivative
$\partial_{;s} a_{s,i_1 \ldots i_l,j_1 \ldots j_k}$ is again
symmetric in indices $(i_1, \ldots, i_l)$ and  antisymmetric in
$(j_1, \ldots, j_k)$. It means that the operation $\partial_{;s}$
changes neither the degree of the Weyl algebra element nor  the
degree of the form. The new index `$s$'  does not generate inner
symmetries in $\partial_{;s} a_{t,i_1 \ldots i_l,j_1 \ldots j_k}$.
The covariant derivative is non symmetric in $(s,i_1, \ldots,
i_l)$ and non antisymmetric in $(s,j_1, \ldots, j_k)$. The
straightforward consequence of this fact is that $\partial_{;s}
a_{t,i_1 \ldots i_l,j_1 \ldots j_k}$ is not an element of ${\cal
P^*M}[[\hbar]] \Lambda$.

Such a result is not compatible with our expectations. Since the elements
$\partial_{;s} a_{k,i_1 \ldots i_l,j_1 \ldots j_k}$ are forms with values in the Weyl algebra we
look for a linear operator which  transforms a $k$-form into a $(k+1)$-form. 
This condition is
fulfilled by the exterior covariant derivative operator \cite{poor}.
\begin{de}
\label{potrzeba}
The exterior covariant derivative 
$\partial:{\cal P^*M}[[\hbar]]\otimes  \Lambda^k \rightarrow {\cal P^*M}[[\hbar]]\otimes 
\Lambda^{k+1}$ is a linear differential operator such that for every $a \in {\cal P^*M}[[\hbar]] 
\Lambda$
\be
\label{x2}
\partial a :=  dq^s \wedge \partial_{;s}a.
\ee
\end{de}
Note that in Fedosov's publications \cite{6,7} the  operator $\partial$ is simply called
`connection'.

\begin{tw} \cite{7}
The exterior covariant derivative $\partial $ in the bundle ${\cal P^*M}[[\hbar]] \Lambda$
in a Darboux atlas $\{(U_{\varrho},\varphi_{\varrho})\}_{\varrho
\in I}$ can be written as
\be
\label{kon1}
\partial a = da + \frac{1}{i \hbar}[\Gamma,a],
\ee
where $a \in {\cal P^*M}[[\hbar]] \Lambda$ and
\be
\label{kon2}
\Gamma :=  \frac{1}{2}\Gamma_{ij,k} dq^k
\ee
is a $1$-form symmetric in $(i,j,k)$.
\end{tw}

It is worthy to make the following remarks: \begin{itemize}
 \item
The formula (\ref{kon1}) is  valid only in Darboux coordinates,
because only in such charts
  $\Gamma_{ijk}$ is symmetric in all the indices.

\item
We consider  $\Gamma$ like an element of ${\cal
P^*M}[[\hbar]] \otimes  \Lambda^1$ although we know that the
elements $\Gamma_{ijk}$ are not components of any tensor. This
fact does not influence on the $\circ$-product in a Darboux chart.

\item 
The formula (\ref{kon1}) is similar to the definition of the
connection matrix \cite{3} but, in fact, the $1$-form (\ref{kon2})
is not the connection matrix. In the analyzed case the connection
matrix  is infinite dimensional.
\end{itemize}

\begin{tw} \cite{7}
The exterior covariant derivative preserves the degree of the forms of
${\cal P^*M}[[\hbar]] \Lambda$.
\end{tw}
\noindent \underline{Proof}.- The derivative $d$ does not change
the degree of the forms of
${\cal P^*M}[[\hbar]] \Lambda$. The commutator increases the power of $\hbar$
in one unit since it has two derivatives  in $X^i$'s, but the fact of
 dividing  by $i\hbar$ finally preserves  the degree.
\hfill\rule{2mm}{2mm}

\begin{tw}\label{teorema4}
For every two forms $ a \in {\cal P^*M}[[\hbar]] \otimes \Lambda^{k_1}$ and $b \in {\cal
P^*M}[[\hbar]]
\otimes \Lambda^{k_2}$ we have
\be
\label{komu6}
\partial (a \circ b) = \partial a \circ b + (-1)^{k_1}a \circ \partial b.
\ee
\end{tw}
\noindent \underline{Proof}.- We make the proof in Darboux
coordinates.
\[\begin{array}{l}
\partial (a \circ b) :=  d (a \circ b) +\frac{1}{i \hbar}[\Gamma, a \circ b] \\[0.25cm]
\qquad\stackrel{\rm  (\ref{komu2})}{=}
da \circ b + (-1)^{k_1} a \circ db
+\frac{1}{i \hbar}(-1)^{1 \cdot k_1}a \circ [\Gamma,   b]
+ \frac{1}{i \hbar}  [\Gamma,   a] \circ
b =\partial a \circ b + (-1)^{k_1}a \circ \partial b.
\end{array}\]
\ \hfill \rule{2mm}{2mm}

Note that the relation (\ref{komu6}) is a Leibniz rule for the exterior covariant derivative
$\partial$ and the $\circ$-product.

Let us consider now the second exterior covariant derivative $\partial (\partial a)$.
As before we make our computations in a Darboux chart. Thus,
\be
\label{i}
\partial (\partial a) = \partial \left( da + \frac{1}{i \hbar}[\Gamma,a]
\right) = d(da) + d \left( \frac{1}{i \hbar}[\Gamma,a]
\right) + \frac{1}{i \hbar}[\Gamma,da + \frac{1}{i \hbar}[\Gamma,a]].
\ee
Using
\[
d (a \wedge b) = da \wedge b + (-1)^{k}a \wedge db, \qquad a \in \Lambda^{k},
\]
which is a general property valid for any  form, and remembering that $\Gamma$ is a $1$-form, we
obtain
\be\begin{array}{l}
\label{kom10}
d(\Gamma \circ a ) = d \Gamma \circ a - \Gamma \circ da
\\[0.25cm]
d( a \circ \Gamma ) = d a \circ \Gamma + (-1)^k a \circ d \Gamma.
\end{array}\ee
From (\ref{kom}) and (\ref{kom10})   we obtain
\be
\label{kom12}
d [\Gamma,a] = [d \Gamma,a] - [\Gamma, da].
\ee
From the Jacobi identity (\ref{kom01})
\be
\label{kom13}
(-1)^{1 \cdot (1 +k)}[\Gamma,[\Gamma,a]] +
 (-1)^{1 \cdot (1 +k)}[a,[\Gamma,\Gamma]] +  (-1)^{k \cdot (1 + 1)}[\Gamma,[a,\Gamma]] =0
\ee
and using the relation
\be
[\Gamma,[a,\Gamma]] \stackrel{\rm (\ref{kom1})}{=} (-1)^{1 \cdot (1 +k)} [\Gamma,[\Gamma,a]]
\ee
we see that
\be
\label{kom14}
2[\Gamma,[\Gamma,a]] = - [a,[\Gamma,\Gamma]] \stackrel{\rm (\ref{kom1})}= [[\Gamma,\Gamma],a].
\ee
Moreover,
\be
\label{kom20}
[\Gamma,\Gamma] = 2 \Gamma \circ \Gamma.
\ee
Putting (\ref{kom12}) and (\ref{kom14}) with (\ref{kom20}) in (\ref{i}) we get
\be
\label{kom15}
\partial (\partial a) = \frac{1}{i \hbar}[d \Gamma + \frac{1}{i \hbar}\Gamma \circ \Gamma,a].
\ee We see that the $2$-form 
\be \label{curvatura}
R= \Gamma + \frac{1}{i \hbar} \Gamma
\circ \Gamma
\ee
 is  the curvature of the connection $\Gamma$. In Darboux charts we can write \be
\label{kom16}
\partial (\partial a) = \frac{1}{i \hbar}[R,a].
\ee
Let us find the explicit form of the $2$-form of curvature (\ref{curvatura}).
For simplicity we consider elements of the Weyl algebra bundle ${\cal P^*M}[[\hbar]]$ acting on
vectors. By $X$ we denote an arbitrary fixed  vector field on  $M$.   
\be\begin{array}{l}\label{kom17}
d\Gamma(X,X) = d\left( \frac{1}{2}\Gamma_{ijk}X^iX^jdq^k \right)
=\frac{1}{2} \frac{\partial \Gamma_{ijk} }{\partial q^l}X^iX^jdq^l \wedge dq^k \\[0.25cm]
\qquad = \frac{1}{4}\left(\frac{\partial \Gamma_{ijk}}{\partial q^l} - \frac{\partial
\Gamma_{ijl}}{\partial q^k}
\right) X^iX^jdq^l \wedge dq^k
=\frac{1}{4} \omega_{is}\left(\frac{\partial \Gamma^s_{kj}}{\partial q^l} - \frac{\partial
\Gamma^s_{lj}}{\partial q^k}
\right) X^iX^jdq^l \wedge dq^k ,
\end{array}\ee
and
\be\begin{array}{l}\label{kom18}
\Gamma \circ \Gamma (X,X)  :=  \frac{i \hbar}{2} \omega^{i_1 j_1}
\cdot 2 \cdot 2 \cdot \frac{1}{2}\Gamma_{i_1il}X^idq^l \wedge
\frac{1}{2}\Gamma_{j_1jk}X^jdq^k \\[0.25cm]
\qquad = \frac{i \hbar}{2} \omega^{i_1 j_1}
\omega_{i_1 u} \Gamma^{u}_{il} \Gamma_{j_1jk}
 X^i X^j dq^l \wedge dq^k
= - \frac{i \hbar}{2}
\Gamma^{j_1}_{il}\Gamma_{j_1jk}
 X^i X^j dq^l \wedge dq^k \\[0.25cm]
\qquad = -\frac{i \hbar}{2} \omega_{js}
\Gamma^{u}_{il}\Gamma^s_{uk}
 X^i X^j dq^l \wedge dq^k
= \frac{i \hbar}{4} \omega_{is} \left(
\Gamma^{u}_{jk}\Gamma^s_{ul} - \Gamma^{u}_{jl}\Gamma^s_{uk} \right)
 X^i X^j dq^l \wedge dq^k.
\end{array}\ee
From (\ref{kom17}) and the last expression of (\ref{kom18}) we obtain that
\be
\label{inny}
R(X,X)= \frac{1}{4} R_{ijlk}  X^i X^j dq^l \wedge dq^k
\ee
where
\be
 R_{ijlk}=\omega_{si}\left( \frac{\partial \Gamma^s_{kj}}{\partial q^l}
 - \frac{\partial \Gamma^s_{lj}}{\partial q^k}
 +\Gamma^{u}_{jk}\Gamma^s_{ul} -  \Gamma^s_{uk} \Gamma^{u}_{jl} \right)
\ee
is the curvature tensor (\ref{now9.4}) of the symplectic connection $\Gamma_{ijk} $ on the phase
space $ M$.
From (\ref{inny}) we can see easily that $R_{ijlk}$ is symmetric in the indices $\{i,j\}$ and
antisymmetric in $\{k,l\}$.

 It is can be  proved that  although we work in a Darboux chart, the relations
(\ref{inny}) and (\ref{kom16}) hold in every chart on ${\cal M}$.

Let us introduce two  operators acting on the bundle ${\cal P^*M}[[\hbar]]\Lambda$. To make the
notation more clear we will operate on elements of the Weyl algebra acting on
vectors of the tangent space $TM$ to the symplectic  manifold  $M $, i.e.
$a_{i_1, \ldots, i_l,j_1,
\ldots, j_s  }dq^{j_1} \wedge \cdots \wedge dq^{j_s}(X^1,\ldots,X^l)$.
\begin{de}
The operator
$\delta:{\cal P^*M}[[\hbar]] \otimes \Lambda^s \rightarrow {\cal P^*M}[[\hbar]] \otimes
\Lambda^{s+1}$    defined as
\be
\label{delta1}
\delta a :=  dq^k  \wedge (\frac{\partial a}{\partial X^k})
\ee
 is known as   {antiderivation}.
\end{de}
In order to see if the above definition is consistent, let us compute the antiderivation of the
constant tensor field.
From the equation (\ref{delta1}) we obtain
\[\begin{array}{l}
\delta (1_{i_1, \ldots, i_l,j_1, \ldots, j_s  }dq^{j_1} \wedge \cdots \wedge dq^{j_s})\\[0,25cm]
\qquad = X^{i_2}\ldots
X^{i_l} dq^{i_1} \wedge dq^{j_1} \wedge \ldots \wedge dq^{j_s} + X^{i_1} X^{i_3}\ldots X^{i_l}
dq^{i_2} \wedge dq^{j_1} \wedge \ldots \wedge dq^{j_s}
\\[0,25cm]
\qquad\qquad + \cdots + X^{i_1}\ldots X^{i_{l-1}} dq^{i_{l-1}} \wedge dq^{j_1} \wedge \ldots \wedge
dq^{j_s} .
\end{array}\]
Hence
\[\begin{array}{l}
\delta (1_{i_1, \ldots, i_l,j_1, \ldots, j_s  }dq^{j_1} \wedge \cdots
\wedge dq^{j_s})= 1_{i_2, \ldots, i_l,j_1, \ldots, j_s  }dq^{i_1} \wedge dq^{j_1} \wedge \cdots
\wedge dq^{j_s}
\\[0.25cm]
\qquad + 1_{i_1,i_3, \ldots, i_l,j_1, \ldots, j_s  }dq^{i_2} \wedge dq^{j_1} \wedge \cdots \wedge
dq^{j_s} + \cdots + 1_{i_1, \ldots, i_{l-1},j_1, \ldots, j_s  }dq^{i_l} \wedge dq^{j_1} \wedge
\cdots \wedge dq^{j_s},
\end{array}\]
which, of course, belongs to ${\cal P^*M}[[\hbar]] \otimes \Lambda^{s+1}$.

\begin{tw}
The operator $\delta$ may be written  as a $1$-form
\be
\label{brak}
\delta = -\frac{1}{i \hbar} [\omega_{ij}X^idq^j, \cdot].
\ee
\end{tw}

\begin{tw}
The operator $\delta$ lowers the degree of the  elements of  
${\cal P^*M}[[\hbar]]  \Lambda$ of $1$.
\end{tw}
The proofs of both theorems  are straightforward.

\begin{de}
The operator $\delta^*:{\cal P^*M}[[\hbar]] \otimes \Lambda^s
\rightarrow {\cal P^*M}[[\hbar]] \otimes \Lambda^{s-1}$ is defined
as \be \label{delta02} \delta^* a :=  X^k \left( \frac{\partial
}{\partial q^k}\right) \rfloor a. \ee
\end{de}
It can be considered as the `opposite' of the antiderivation
operator. The symbol $\rfloor $ denotes the inner product.
\noindent Effectively, from the definition (\ref{delta02}) we
obtain that \be\begin{array}{l} \label{cosik1}
\delta ^*(X^{i_1}\ldots X^{i_l} dq^{j_1} \wedge \ldots \wedge dq^{j_s})=\\[0.25cm]
\qquad X^{j_1}X^{i_1}\ldots
X^{i_l} dq^{j_2} \wedge \ldots \wedge dq^{j_s}  - X^{j_2}X^{i_1}\ldots X^{i_l} dq^{j_1} \wedge
dq^{j_3} \wedge \ldots \wedge dq^{j_s}\\[0.25cm]
\qquad\qquad+ \cdots + (-1)^{s+1} X^{j_s}X^{i_1}\ldots X^{i_l} dq^{j_1}
\wedge \ldots \wedge dq^{j_{s-1}}.
\end{array}\ee

\begin{tw} \cite{7}
The operator $ \delta^*$ raises the degree of the forms of ${\cal P^*M}[[\hbar]] \Lambda$ in
$1$.
\end{tw}
\noindent \noindent \underline{Proof}.- The operator $ \delta^*$
exchanges the $1$-form $dq^k$ into $X^k$ and does not touch
$\hbar$, so the degree increases in $1$. \hfill\rule{2mm}{2mm}

\begin{tw} \cite{7}
\label{nieoczekiwane}
The operators $\delta$ and $\delta^*$ do not depend on
the choice of local coordinates and have the following properties:

(i) linearity,

(ii)
\[
\delta^2 = (\delta^*)^2=0,
\]

(iii)
for the monomial $ X^{i_1}\ldots X^{i_l} dq^{j_1} \wedge \ldots \wedge dq^{j_s}$ we have
\[
(\delta \delta^* + \delta^* \delta)X^{i_1}\ldots X^{i_l} dq^{j_1}
\wedge \ldots \wedge dq^{j_s} = (l+s) X^{i_1}\ldots X^{i_l}
dq^{j_1} \wedge \ldots \wedge dq^{j_s}.
\]
\end{tw}
\noindent
\underline{Proof}.- Linearity is obvious, thus let us start from
the second property. From the symmetry of $ X^{i_1}\ldots X^{i_l}
dq^{j_1} \wedge \ldots \wedge dq^{j_s}$ in $X^i$'s it is
sufficient to compute $\delta^2$
 for two fixed $X$'s. Hence
 \[
 \delta^2(X^{i_1}X^{i_2}) = \delta (X^{i_2}dq^{i_1} + X^{i_1}dq^{i_2})=dq^{i_2} \wedge  dq^{i_1} + dq^{i_1} \wedge dq^{i_2} =0.
\]
Now from the antisymmetry of $ X^{i_1}\ldots X^{i_l} dq^{j_1} \wedge \ldots \wedge dq^{j_s}$ in
$dq^{j}$'s  it is enough to find $ (\delta^*)^2 (dq^{j_1} \wedge
dq^{j_2})$  for two fixed $dq^{j_1}$ and $dq^{j_2}$. So,
\[
(\delta^*)^2(dq^{j_1} \wedge dq^{j_2}) = \delta^* (X^{j_1}dq^{j_2}- X^{j_2}dq^{j_1})= X^{j_1}X^{j_2} - X^{j_2}X^{j_1}=0.
\]
To show the third statement we consider the monomial $X^idq^j$. The result is easy to generalize on  
$ X^{i_1}\ldots X^{i_l} dq^{j_1} \wedge \ldots \wedge dq^{j_s}$ using the symmetry in $X^i$'s and
the antisymmetry in  $dq^j$'s. Now
\[
(\delta \delta^* + \delta^* \delta)(X^idq^j)=\delta(X^iX^j)
+ \delta^*(dq^i \wedge dq^j) = X^i dq^j + X^j dq^i + X^i dq^j - X^j dq^i = 2 X^i dq^j.
\]
\hfill\rule{2mm}{2mm}
\begin{de} \cite{7}
There is an operator $\delta^{-1}:{\cal P^*M}[[\hbar]] \otimes \Lambda^s \rightarrow {\cal
P^*M}[[\hbar]]
\otimes \Lambda^{s-1}$  defined by
\be
\label{delta2}
\delta^{-1} a := \left\{ \begin{array}{ccl}
&\frac{1}{l+s} \delta^* a \qquad  &{\rm for} \;\;\; l+s>0   \\[0.35cm]
 & 0 \qquad &{\rm for} \;\;\; l+s=0
\end{array}\right.
\ee
where $l$ is  the degree of $a$ in $X^i$'s (i.e. the number of $X$'s) and $s$ is the degree
of the form.
\end{de}
\noindent
Note that, in fact, $\delta^{-1}$ is, up to constant, the $\delta^*$ operator.

The straightforward consequence of the linearity and the decomposition on monomials is the de Rham
decomposition of the form $a \in {\cal P^*M}[[\hbar]]\Lambda$ as shows next theorem.
\begin{tw} \cite{7}
For every $a \in {\cal P^*M}[[\hbar]]\Lambda$ the following equality holds
\be
\label{dec1}
a= \delta \delta^{-1}a + \delta^{-1} \delta a +a_{00},
\ee
where $ a_{00} $ is a function on the symplectic manifold $M$.
\end{tw}
\noindent \underline{Proof}.- For functions $a_{00}$ on  $M$ the
fact that the relation (\ref{dec1}) holds is evident. To show that
the de Rham decomposition is true also for elements of ${\cal
P^*M}[[\hbar]]\Lambda$ we use the fact that the operators $\delta$ and
$\delta^{-1}$ are linear. From the theorem (\ref{nieoczekiwane})
we have   for a monomial  that \be \label{poco}
 (\delta \delta^* + \delta^* \delta) (X^{i_1}\ldots X^{i_l} dq^{j_1} \wedge \ldots \wedge dq^{j_s})=
(l+s) X^{i_1}\ldots X^{i_l} dq^{j_1} \wedge \ldots \wedge dq^{j_s}.
\ee
From the definition (\ref{delta2}) the l.h.s. of the above equation may be written as
\[
\left(\delta (l+s)\delta^{-1} + ((l+1)+(s-1))\delta^{-1}
\delta \right) X^{i_1}\ldots X^{i_l} dq^{j_1} \wedge \ldots \wedge dq^{j_s} .
\]
Hence,  from (\ref{poco}) we immediately obtain (\ref{dec1}).
\hfill\rule{2mm}{2mm}

 The definition (\ref{potrzeba}) of the exterior
covariant derivative is based on covariant derivatives determined
by the symplectic connection. It is possible to generalize this
description of exterior covariant derivative (compare with \cite{se1}
where the term `connection' is used instead of
`exterior covariant derivative').
\begin{de}
The exterior covariant derivative 
$\tilde{\partial}:{\cal P^*M}[[\hbar]]\otimes  \Lambda^0 \rightarrow {\cal P^*M}[[\hbar]]\otimes 
\Lambda^{1}$ is a linear differential operator such that for every $a \in {\cal P^*M}[[\hbar]] 
\otimes \Lambda^0$ and $f \in \Lambda^{0}$
\be
\label{cos1.2}
\tilde{\partial}(f \cdot a) = df \otimes a + f \cdot \tilde{\partial}a.
\ee
\end{de}
This definition may be extended to an arbitrary $k$-form with values in the Weyl algebra ${\cal
P^*M}[[\hbar]]$.
\begin{tw} \cite{se1}
\label{seba1}
There is a unique operator 
$\breve{\partial}: {\cal P^*M}[[\hbar]]\otimes  \Lambda^k \rightarrow {\cal P^*M}[[\hbar]]\otimes 
\Lambda^{k+1}$ satisfying:
\begin{enumerate}
\item
\be
\label{se5}
\breve{\partial}(f \wedge a)= df \wedge a + (-1)^s f \wedge \breve{\partial}a
\ee
for every $f \in \Lambda^s, \;\; a \in {\cal P^*M}[[\hbar]]\otimes  \Lambda^k$.  
\item
\be\label{conexion}
\breve{\partial} a = \tilde{\partial} a
\ee
for $a \in {\cal P^*M}[[\hbar]]\otimes  \Lambda^0$.
\end{enumerate}
\end{tw}

In a local Darboux chart we can write
\be
\label{cos2}
\breve{\partial}a = da + \frac{1}{i \hbar}[\Phi , a],
\ee
where $\Phi \in {\cal P^*M}[[\hbar]]\otimes  \Lambda^1$ 
is a connection determining the derivation $\breve{\partial}$. The same definition works for
$0$-forms, so if $a \in {\cal P^*M}[[\hbar]]\otimes  \Lambda^{0}$
\be
\label{cos222}
\tilde{\partial}a = da + \frac{1}{i \hbar}[\Phi , a].
\ee
\newline
Indeed, the operator $d + \frac{1}{i \hbar}[\Phi , \cdot]$ is linear. For every $a \in {\cal P^*M}[[\hbar]]  \Lambda^0$ and $f \in \Lambda^{0}$
\[
\tilde{\partial}a=
d(f \cdot a) + \frac{1}{i \hbar}[\Phi , f \cdot a]= df \cdot a + (-1)^s f \wedge da + f \cdot \frac{1}{i \hbar}[\Phi ,  \cdot a ]=
df \cdot a + f \cdot \tilde{\partial}a .
\]
Hence, the condition (\ref{cos1.2}) is fulfilled.
Now for $a \in {\cal P^*M}[[\hbar]]  \Lambda^k$ and $f \in \Lambda^{s}$ we
see   from (\ref{cos2}) that the property (\ref{se5}) holds, i.e.
\[\begin{array}{ccl}
\breve{\partial}({f \wedge a}) &=&
d(f \wedge a) + \frac{1}{i \hbar}[\Phi , f \wedge a] \\[0.25cm]
 &\stackrel{\;\rm (\ref{kom})}{=}&
df \wedge a + (-1)^s f \wedge da +  \frac{1}{i \hbar} \Phi \circ (f \wedge a)  
- (-1)^{1 \cdot (s+k)}  \frac{1}{i \hbar} (f \wedge a) \circ \Phi \\[0.25cm] 
&=& df \wedge a + (-1)^s f \wedge da 
+ (-1)^s f \wedge \frac{1}{i \hbar} \Phi \circ a
-(-1)^s f \wedge (-1)^k \frac{1}{i \hbar} a \circ \Phi \\[0.25cm]
&=& df \wedge a + (-1)^s  f \wedge
\left( da +
\frac{1}{i \hbar}[\Phi,a]
\right)\\[0.25cm]
&=&df \wedge a + (-1)^s f \wedge \breve {\partial}a .
\end{array}\]
Since  $\tilde{\partial} a =\breve{\partial}a$
 for every $ a \in {\cal P^*M}[[\hbar]] \otimes  \Lambda^0$ we can use
  the formula (\ref{cos2}) as   definition of the exterior covariant derivative in 
${\cal P^*M}[[\hbar]]   \Lambda$.

 Analogously to the case of the connection $\Gamma$ (\ref{curvatura}) we define the curvature of
$\Phi$ as a $2$-form with values in the Weyl algebra
\be
\label{cos3}
\breve{R}= d \Phi + \frac{1}{i \hbar} \Phi \circ \Phi.
\ee
Of course, in  Darboux coordinates for every $a \in {\cal P^*M}[[\hbar]]  \Lambda $ we can write
\be \label{dda}
\breve{\partial}(\breve{\partial}a)=  \frac{1}{i \hbar}[\breve{R},a].
\ee
Among the properties of the curvature $\breve{R}$ we  present the so-called
 Bianchi identity.
\begin{tw}
For the curvature $\breve{R}$ the following relation holds
\be
\label{bianchi}
\breve{\partial}\breve{R}=0.
\ee
\end{tw}
\noindent
\underline{Proof}.-
\newline
\[\begin{array}{l}
\breve{\partial}\breve{R} \stackrel{\rm (\ref{cos2})}{=}d \breve{R} 
+ \frac{1}{i \hbar}[\Phi, \breve{R}]
\stackrel{\rm (\ref{cos3})}{=} 
d \left( d \Phi + \frac{1}{i \hbar} \Phi \circ \Phi \right) + \frac{1}{i \hbar}
[\Phi, d \Phi + \frac{1}{i \hbar} \Phi \circ \Phi] \\[0.25cm]
\qquad \; = d^2 \Phi + \frac{1}{i \hbar} d \Phi \circ \Phi + (-1)^1 \frac{1}{i \hbar} \Phi \circ d
\Phi + \frac{1}{i \hbar} \Phi \circ d \Phi 
- \frac{1}{i \hbar}(-1)^2 d \Phi \circ \Phi + \frac{1}{(i \hbar)^2}[\Phi, \Phi \circ \Phi]
\\[0.25cm]
\qquad \; = \frac{1}{(i \hbar)^2} \left( \Phi \circ \Phi \circ \Phi 
- (-1)^{1 \cdot 2} \Phi \circ \Phi \circ \Phi \right)=0.
\end{array}\]
\hfill \rule{2mm}{2mm}

\vspace{0.5cm}
 In our considerations a special role will be played by the so called `abelian connection'.

\begin{de} \cite{7}
A connection $D$ (defined by (\ref{conexion})) in the bundle ${\cal P^*M}[[\hbar]]  \Lambda$ is
called {abelian} if for any section $a \in C^{\infty}({\cal P^*M}[[\hbar]]  \Lambda) $
\be
D^2 a =\frac{1}{i \hbar} [\Omega,a]=0.
\ee
\end{de}
The curvature $\Omega$ of the abelian connection $D$ is a central form. 
Since $\Omega$ is central we deduce that
\be
d \Omega=0.
\ee
Indeed, from the general formula (\ref{cos2}) we have
\[
D \Omega = d \Omega + \frac{1}{i \hbar}[\tilde{\Gamma},\Omega],
\]
where by $\tilde{\Gamma}$ we denote the $1$-form of the abelian connection. 
Since $\Omega$ is a central form   the commutator
$[\tilde{\Gamma},\Omega]$ disappears. From the other side, the Bianchi identity (\ref{bianchi})
holds for the curvature  $\Omega$, i.e.  
$ D \Omega =0 $. Hence  $ d \Omega =0 $.

Let us assume that in a Darboux chart
\be
\label{koo1}
D = \partial +\frac{1}{i \hbar}[(\omega_{ij}X^idq^j + r),\cdot],
\ee
where $r$ is a globally defined $1$-form satisfying the Weyl normalizing condition, i.e. the part of
$r$ not containing $X^i$'s vanishes. Let us compute the curvature of the connection (\ref{koo1}). As
we have mentioned before in (\ref{cos3})
\[
\Omega = d\tilde \Gamma + \frac{1}{i \hbar}\tilde{\Gamma} \circ  \tilde{\Gamma},
\]
where
\be\label{nieoc}
\tilde{\Gamma}:=  \Gamma + \omega_{ij}X^idq^j + r.
\ee
We can see that
\be\begin{array}{l}\label{koo0}
\Omega = d \Gamma + dr + \frac{1}{i \hbar} \Gamma \circ \Gamma 
+\frac{1}{i \hbar} \Gamma \circ \omega_{ij}X^idq^j + \frac{1}{i \hbar} \Gamma \circ r
 +\frac{1}{i \hbar} \omega_{ij}X^idq^j \circ \Gamma \\[0.25cm]
\;\; + \frac{1}{i \hbar} \omega_{ij}X^idq^j \circ \omega_{kl}X^kdq^l
+ \frac{1}{i \hbar} \omega_{ij}X^idq^j \circ r + \frac{1}{i \hbar} r \circ \Gamma  +
\frac{1}{i\hbar} r \circ \omega_{ij}X^idq^j
 + \frac{1}{i \hbar} r \circ r.
\end{array}\ee
 We know that 
\be \label{koo2} 
d \Gamma  + \frac{1}{i \hbar}
\Gamma \circ \Gamma =R. 
\ee 
Since our definition has been formulated in
a Darboux chart we can write
 \be \label{koo3} 
dr + \frac{1}{i \hbar} \Gamma
\circ r  + \frac{1}{i \hbar} r \circ \Gamma = dr + \frac{1}{i
\hbar}[\Gamma,r]= \partial r.
\ee 
Moreover, \be \label{koo4}
\frac{1}{i \hbar} \Gamma \circ \omega_{ij}X^idq^j  +\frac{1}{i
\hbar} \omega_{ij}X^idq^j \circ \Gamma = \frac{1}{i \hbar}[\Gamma,
\omega_{ij}X^idq^j]= \frac{1}{2} \cdot 2 \cdot 2 \cdot
\frac{1}{2}\omega^{li}\Gamma_{lrs}X^r dq^s  \wedge
\omega_{ij}dq^j. 
\ee 
We know that (\ref{now3}) holds, hence
(\ref{koo4}) equals to
 \be \label{koo5} 
\delta^l_j \Gamma_{lrs}X^r
dq^s \wedge dq^j= \Gamma_{lrs}X^r dq^s \wedge dq^l=0, 
\ee 
because
$ \Gamma_{lrs}$ is symmetric in all the indices. Now 
\be \label{koo6}
\frac{1}{i \hbar} \omega_{ij}X^idq^j \circ r + \frac{1}{i \hbar} r
\circ \omega_{ij}X^idq^j = \frac{1}{i
\hbar}[\omega_{ij}X^idq^j,r]\stackrel{\rm (\ref{brak})}{=}- \delta
r. 
\ee
 Finally \be \label{koo7} \frac{1}{i \hbar}
\omega_{ij}X^idq^j \circ \omega_{kl}X^kdq^l=
-\frac{1}{2}\omega_{ij}dq^i \wedge dq^j. 
\ee 
Putting (\ref{koo2})--(\ref{koo7}) into (\ref{koo0}) we obtain 
\be \label{koo8} 
\Omega
= -\frac{1}{2}\omega_{ij}dq^i \wedge dq^j +R -\delta r + \partial
r + \frac{1}{i \hbar} r \circ r. 
\ee
The abelian property will be
fulfilled, provided \be \label{koo9} \delta r= R  + \partial r +
\frac{1}{i \hbar} r \circ r. 
\ee

\begin{tw} \cite{7}
The equation (\ref{koo9})
has a unique solution satisfying the following conditions
\be
{\rm deg} \; r \geq 3, \qquad \delta^{-1}r=0.
\ee
\end{tw}
\noindent
\underline{Proof}.- 
From the decomposition (\ref{dec1}) we obtain
\[
r=\delta^{-1} \delta r,
\]
because $r$ is a $1$-form we have  $r_{00}=0$ and $\delta^{-1}r=0$. 
Moreover, $\delta r$ is a solution of (\ref{koo9}). Hence
\be
\label{koo10}
r=\delta^{-1} (R  + \partial r + \frac{1}{i \hbar} r \circ r).
\ee
The operator $\delta^{-1}$ raises the degree by $1$, so (\ref{koo10}) 
is a recurrent formula starting with $ \delta^{-1} R$.
The proof that solution (\ref{koo10}) really fulfills (\ref{koo9}) and that it is unique, 
is more complicated (see it in \cite{7} or \cite{12}).
\hfill\rule{2mm}{2mm}

Note that  we have not included in the expression of $r$ terms with degrees lower than $3$. This is
due to the fact that the abelian connection $\tilde{\Gamma}$  (\ref{nieoc}) contains
terms with degrees
$1$ and $2$ and for that  $r$, from its definition, is an object with degree ${\rm deg}\; r \geq 3$.

\begin{tw}
For the abelian connection $D$ and two forms 
$ a \in {\cal P^*M}[[\hbar]]   \otimes \Lambda^{k_1}$ and $ b \in {\cal P^*M}[[\hbar]]   \otimes
\Lambda^{k_2}$ we have
\be
\label{komu9}
D (a \circ b) = D a \circ b + (-1)^{k_1}a \circ D b.
\ee
\end{tw}
The proof is analogous to that of theorem \ref{teorema4}.

\begin{co} \cite{7}
The set of $0$-forms such that their abelian connection vanishes constitutes 
the subalgebra ${\cal P^*M}_{D}[[\hbar]]$ of  ${\cal P^*M}[[\hbar]]$. 
\end{co}

\sect{The $*$-product on $M$}\label{starproduct}

In former sections we have studied the structure of the Weyl
algebra bundle. Thus, we are able to introduce the $*$-product on the
symplectic manifold $M$.  
\begin{de}
The {projection} $\sigma:{\cal P^*M}[[\hbar]] \rightarrow
C^{\infty}(M) $ assigns to each $0$-form, $a$, of ${\cal
P^*M}[[\hbar]]$   its part $a_{00}$ (according to its the de Rham decomposition  
(\ref{dec1})), i.e. 
\be 
\sigma(a) := a_{00}. 
\ee
\end{de}

The following theorem holds.
\begin{tw} \cite{7}
\label{noco} For any function $a_{00} \in C^{\infty}(M)$ there exists a
unique section  $a$ of $C^{\infty}({\cal P^*M}_{D}[[\hbar]]) $ such that $
\sigma(a) = a_{00}$. This element is defined by the recurrent formula
\be \label{fin1} a= a_{00} + \delta^{-1}(\partial a + \frac{1}{i
\hbar}[r,a]). 
\ee
\end{tw}
\noindent The recurrent character of that solution  can be proved analogously to
(\ref{koo10}) (for more details see  \cite{7}).

The map $ \sigma$ gives a one-to-one correspondence between 
$C^{\infty}({\cal P^*M}_{D}[[\hbar]]) $ and $ C^{\infty}(M)$. Then,  
$a =\sigma^{-1}(a_{00})$, where $\sigma^{-1}$ is the inverse map of $\sigma $.  This 
map  provides a quantization procedure as follows. 

\begin{de} \cite{7}
Let $F_1, F_2$ be two  $ C^{\infty}(M)$-functions. 
The $*$-{product} is defined as \be \label{koniec} F_1 * F_2 :=
\sigma\left( \sigma^{-1}(F_1) \circ \sigma^{-1}(F_2) \right). \ee
\end{de}
This $*$-product can be considered  as a generalization of the
Moyal product of  Weyl type defined for the case $ M = {\mathbb
R}^{2n}$. It has the following properties:
\renewcommand{\theenumi}{\bf \arabic{enumi}}
\begin{enumerate}
\item Invariance under Darboux transformations. In fact, the
$*$-product is invariant under all  smooth transformation of
coordinates on $M$. However, the  definitions of exterior
covariant derivatives $ \partial$ and $D$ cannot be written in the
form (\ref{kon1}) and (\ref{koo1}), respectively.

\item In the limit $\hbar \rightarrow 0^+ $ the $*$-product turns
into the commutative point-wise multiplication of functions, i.e.
\be\label{limite} \lim_{\hbar \rightarrow 0^+ }F_1 * F_2= F_1
\cdot F_2. \ee
 Indeed, since  in (\ref{koniec})
 only terms not containing $X^i$'s and $\hbar^n, \; n \geq 1$, 
are  taking  we can see that, in
fact, only the terms of  degree zero  are essential  in the product
$\sigma^{-1}(F_1) \circ \sigma^{-1}(F_2) $. From the third
property of the $\circ$-product (\ref{degree}) we
deduce that the part of $\sigma^{-1}(F_1) \circ \sigma^{-1}(F_2) $
with degree zero is just $F_1 \cdot F_2 $. Hence,  we obtain
(\ref{limite}). \item
 The multiplication (\ref{koniec}) is associative but noncommutative.

Effectively, from the theorem (\ref{noco}) the relation
\[
\sigma^{-1}(F_1)=\sigma^{-1}(F_2) \Longleftrightarrow F_1=F_2 
\]
holds. It means that 
\be\label{associativity}
\sigma^{-1}\left((F_1 * F_2)*F_3\right)
{=} \sigma^{-1}(F_1 * F_2)\circ \sigma^{-1}(F_3)
{=} \left(\sigma^{-1}(F_1) \circ
\sigma^{-1}(F_2)\right)\circ \sigma^{-1}(F_3). 
\ee 
Since the
$\circ$-product is associative the r.h.s. of (\ref{associativity})
can be written as
\[
\sigma^{-1}(F_1) \circ \left( \sigma^{-1}(F_2)\circ
\sigma^{-1}(F_3) \right)
 {=}
\sigma^{-1}(F_1) \circ \sigma^{-1}\left((F_2 *F_3)\right)
{=} \sigma^{-1}\left((F_1 *
(F_2*F_3)\right) .
\]
Hence
\[
(F_1 * F_2)*F_3=F_1 * (F_2*F_3), \qquad \forall F_1, F_2, F_3 \in
C^{\infty}(M) .
\]
\item In the case $  M = {\mathbb R}^{2n}$ the product defined
above is just the Moyal product of
 Weyl type.
Instead of proving this statement here we will analyze the case $
M = {\mathbb R}^{2n}$ in the next section.
\end{enumerate}

The properties mentioned above show that the $*$-product
constructed according to the Fedosov idea is a natural
generalization of the $*$-product in the trivial case when $  M =
{\mathbb R}^{2n}$ (Moyal-product). Several  different properties
of the $*$-product may be found in \cite{borm} or \cite{my}.

The construction,  based on fibre  bundle theory, of a $*$-product
of  Weyl type is finished. Remember that giving a symplectic space $M$ equipped
with a symplectic connection a Weyl algebra bundle is constructed.
Operating on  flat sections of this bundle a noncommutative but
associative product of observables on $M$ is defined. It seems to
be the generalization of the Moyal product.

In this paper we have only considered  $*$-products of  the Weyl type. Other
kinds of $*$-products, whose geometrical origin   is the same that
of  the $*$-product of  the Weyl type, were analyzed in \cite{12}.

The existence of many $*$-products on the same symplectic manifold
is closely related to the equivalence problem   of $*$-products.
Two $*$-products $*_1$ and $*_2$ are said to be equivalent iff
there exists  a differential operator $\hat{T}$ such that for
every two functions $F_1, F_2$ of $C^{\infty}(M)$, for which
expressions appearing below have  sense, the following relation
holds \be \label{eqv} F_1 *_1 F_2 = \hat{T}^{-1} \left(\hat{T}F_1
*_2 \hat{T}F_2 \right). \ee It has been  proved in \cite{en1}
that all  the $*$-products on a symplectic manifold are equivalent
to the  Weyl type $*$-product  constructed according to the
Fedosov recipe. For more details about the equivalence problem   of
$*$-products   see \cite{en1} and \cite{en2}.

The definition of the $*$-product is not sufficient to study
completely  most of the  physical problems.  To solve the
eigenvalue equation for an observable $O$ or to find its average
value we must know how to define the states.  There is not a
general answer to this question. It seems that in the framework of
the Fedosov formalism the states are described by  functionals $W$
over some functions defined on the phase space of the system but,
in fact, our knowledge about such objects is rather poor.

Topics connected with the  representation problem of a quantum
state on a phase space such as  `traciality' operation  or
closeness of the $*$-product can be studied, for instance, in
\cite{7,en3,en4}.


\sect{Examples}\label{examples}

In this section we present three systems where the Fedosov
construction is accomplished. The results obtained in two of them
are known from the `traditional' quantum mechanics. The third one,
recently developed by us~\cite{my}, does not have its counterpart
in the formalism of operators in a certain Hilbert space.

\subsect{Cotangent bundle $T^*{\mathbb R}^{n}$ }

In this first example,  we consider the simple case of a physical
system  whose phase space is the vector space $({\mathbb
R}^{2n},\omega)$.  This space is covered with one chart $({\mathbb
R}^{2n}, \varrho),$ in which the symplectic form takes its natural
shape, i.e. \be \label{o0} \omega = \sum_{i=1}^n dq^{n+i} \wedge
dq^i. \ee We choose the symplectic connection in such a way that
all the coefficients vanish, i.e. 
$
\Gamma_{ijk}=0 , \; 1 \leq i,j,k \leq 2n.
$
The symplectic curvature tensor $R_{ijkl}$ also vanishes.

The abelian connection (\ref{nieoc}) in the Weyl bundle  is given
by
\[
\tilde{\Gamma}=  \omega_{ij}X^i dq^j
\]
and its curvature is a central form (\ref{koo8})
\[
\Omega = -\frac{1}{2}\sum_{i=1}^n dq^{n+i} \wedge dq^i .
\]
Hence, $r=0$. It means that for every $F \in C^{\infty}({\mathbb
R}^{2n})$ according to (\ref{fin1}) we can write
\be \label{o1} 
\sigma^{-1}(F)
= F + \delta^{-1}(dF )= \sum_{m=0}^{\infty}\sum_{i_1,i_2, \ldots,
i_m=0}^{m}\frac{1}{m!} \frac{\partial^m F}{\partial q^{i_1}\ldots
\partial q^{i_m}} X^{i_1}\cdots X^{i_m}. 
\ee 
From (\ref{odw2}),
(\ref{6}) and (\ref{o0}) we can see that 
\be \label{o2}
\sigma^{-1}(F_1) \circ \sigma^{-1}(F_2) = F_1(Q + {X}) *_{W}
F_2(Q + {X}), 
\ee 
where $Q=(q^1, \ldots, q^{2n}),\; \;
{X}=(X^1, \ldots, X^{2n}). $ 
The symbol `$*_{W}$'   denotes the
$*$-product of  Weyl type on ${\mathbb R}^{2n}$ defined by the
formula (\ref{odw2}). We can write the equality relation in
(\ref{o2}) because  ${\mathbb R}^{2n}$ and $T {\mathbb R}^{2n}$
are isomorphic.

Now from (\ref{koniec}) we obtain that \be F_1 * F_2 = \sigma
\left(F_1(Q + {X}) *_{W} F_2(Q + {X}) \right)= F_1(Q)
*_{W} F_2(Q). \ee We conclude that  when the symplectic space is
just $({\mathbb R}^{2n},\omega),$ the $*$-product computed using
the Fedosov method is exactly the usual $*$-product of Weyl
type.

\subsect{Harmonic oscillator}

Let us consider a system with phase space $(\mathbb {\mathbb
R}^{2n},\omega)$,
where $\omega = dp \wedge dq$, and   Hamiltonian
\[
H= \frac{p^2}{2m} + \frac{mq^2}{2}.
\]
In a chart with coordinates $(p,q)$ all the coefficients of the
connection $\Gamma_{ijk}, \; i,j,k=1,2$, vanish. In the new
Darboux coordinates $(H, \phi )$, related with the old ones
$(p,q)$    according to 
\be 
q= \sqrt{\frac{2H}{m}} \sin \phi ,
\qquad p= \sqrt{2mH}\cos \phi , 
\ee 
the symplectic form   is rewritten as
 \[
\omega = dH \wedge d \phi. 
\]
 From the
transformation rule for connections (\ref{now7}) we immediately
obtain that the
 symplectic connection $1$-form in coordinates  $(H, \phi)$ is given by
\be \Gamma= \frac{1}{4H}X^2X^2d \phi + \frac{1}{2H}X^1X^2dH + H
X^1X^1dH. \ee Obviously, its curvature vanishes. That implies that
also $r=0$ (see (\ref{koo10})) and the symplectic  curvature in the
Weyl algebra bundle is  central (\ref{koo8})
\[
\Omega = -\frac{1}{2}d H \wedge d \phi,
\]
and the abelian connection is
\[
\tilde{\Gamma}=   X^1 d \phi - X^2 dH + \Gamma.
\]

Let us look for the eigenvalues and eigenfunctions of the
Hamiltonian $H$ in coordinates $(H,\phi)$. The eigenvalue equation
takes the form \be H * W_E(H,\phi)=E \cdot W_E(H,\phi), \ee where
$W_E(H,\phi)$ is  a functional representation of an eigenstate  with
eigenvalue $E$. From (\ref{fin1}) we can write 
\be \label{ha1}
\sigma^{-1}(H) = H + X^2 +H X^1 X^1 + \frac{1}{4H}X^2X^2 
\ee 
and
\[\begin{array}{lll}
\sigma^{-1}(W_E)&=& W_E + \frac{\partial W_E}{\partial \phi}X^1 +
\frac{\partial W_E}{\partial H}X^2 + H \frac{\partial
W_E}{\partial H}X^1 X^1 + \frac{1}{2}\frac{\partial^2
W_E}{\partial \phi^2}X^1X^1 - \frac{1}{2H}\frac{\partial
W_E}{\partial \phi}X^1X^2
\\[0.25cm]
&&\quad + \frac{\partial^2 W_E} {\partial \phi \partial H}X^1X^2 +
\frac{1}{4H}\frac{\partial W_E}{\partial H}X^2X^2 + \frac{1}{2}
\frac{\partial^2 W_E}{\partial H^2}X^2X^2+
  \frac{1}{6}\frac{\partial^3 W_E}{\partial \phi^3}X^1X^1X^1 + \ldots
\end{array}
\]
The series $\sigma^{-1}(W_E)$ is infinite but
only the terms of ${\rm deg}
\leq 2 $ are essential because $\sigma^{-1}(H)$ has degree $2$.
 Computing $\sigma^{-1}(H) \circ
\sigma^{-1}(W_E)$ and projecting the product on the phase space we
finally obtain that
 \be \label{ha3}
H * W_E(H,\phi) = (H-E)W_E - \frac{\hbar^2}{4}\frac{\partial
W_E}{\partial H} - \frac{\hbar^2}{4}H \frac{\partial^2
W_E}{\partial H^2} - \frac{i \hbar}{2}\frac{\partial W_E}{\partial
\phi}  - \frac{\hbar^2}{16H}\frac{\partial^2 W_E}{\partial
\phi^2}=0. \ee Since the function $W_E$ is real we see that 
\be\label{ha4} 
\frac{i \hbar}{2}\frac{\partial W_E}{\partial \phi}=0 .
 \ee 
Thus, $W_E$ depends only on $H$ and  may contain $\hbar$ in
the denominator. It is possible that for some solutions of (\ref{ha3})
the series $\sigma^{-1}(W_E)$ is not well defined. Such functions
are not admissible because for them the product $\sigma^{-1}(H)
\circ \sigma^{-1}(W_E)$ does not exist. However, it happens that
admissible solutions of (\ref{ha3}) are Wigner functions \be
\label{ha5} W_{E_n}(H)= \frac{1}{\pi \hbar}(-1)^n
\exp\left(\frac{-2H}{\hbar}\right)L_n \left(
\frac{4H}{\hbar}\right), \ee where $L_n$ is a Laguerre polynomial,
and the energy $E$ is quantized \be \label{ha6} E_n= \hbar(n +
\frac{1}{2}), \qquad n=0,1,2, \ldots \ee This result is well known
from traditional quantum mechanics. It can be also obtained in
terms of the Moyal product (\ref{odw3})  \cite{cak}.

\subsect{$2$-Dimensional phase space with constant curvature tensor}

Let us consider the phase space $({\mathbb R}^{2},\omega)$.
Topologically this phase space is homeomorphic to ${\mathbb
R}^{2}$. We cover it  with an atlas containing only one chart
$({\mathbb R}^{2n},\varrho). $ In this chart the symplectic form
is, as usual, 
$
\omega = dp \wedge dq,
$
where $q$  denotes the spatial coordinate and $p$ the momentum
conjugate to $q$.
 We assume that in the chart $({\mathbb R}^{2},\varrho)$ the  connection $1$-form is
\be \label{204} \Gamma= \frac{1}{2} p X^1 X^1 dq.
 \ee
The connection (\ref{204}) is well defined
globally since we cover the symplectic manifold $({\mathbb R}^{2},\omega)$ with only
one chart. The curvature 2-form    is 
\be \label{205}
R=-\frac{1}{2}X^1 X^1 dq \wedge dp . 
\ee 
Hence,  in the chart
$({\mathbb R}^{2},\varrho)$ the curvature tensor has only one
nonvanishing component which is  constant \be \label{205a}
 R_{1112}= -R_{1121}= - \frac{1}{4}.
\ee The Ricci tensor is \be \label{205c} K_{11}= \frac{1}{4}. \ee
Since the  connection 1-form is defined by the expression
(\ref{204}), we are able to build the abelian connection. Let us
compute the series $ r $ defined by (\ref{koo10}).  The first term
in the recurrent expression (\ref{koo10}) is
 \be
 \label{206}
\delta^{-1}R = \frac{1}{8} X^1 X^1 X^2 dq - \frac{1}{8} X^1 X^1
X^1 dp. \ee After some computations (for details see \cite{my}) we
obtain that the abelian connection is 
\be\begin{array}{lll}
\label{208}
 \tilde{\Gamma}&=& -X^1dp + X^2 dq + \frac{1}{2} p X^1 X^1 dq
 -\frac{1}{8} X^1 X^1 X^1 dp
   +\frac{1}{8} X^1 X^1 X^2 dq   \nonumber\\[0.25cm]
&& +\frac{1}{128}  X^1 X^1 X^1 X^1 X^1 dp
-\frac{1}{128}  X^1 X^1 X^1 X^1 X^2 dq  \\[0.25cm]
&& - \frac{1}{1024} X^1 X^1 X^1 X^1 X^1 X^1 X^1 dp +
\frac{1}{1024} X^1 X^1 X^1 X^1 X^1 X^1 X^2 dq - \cdots
\end{array}\ee
Note that the abelian connection is an infinite series, the Planck constant
$ \hbar $ does not appear in any term of the series, and
in the series  $ r $ (\ref{nieoc}) only  two kind of terms are present:
\[
(-1)^{m+1} a_{2m+1} (X^1)^{2m}X^2 dq
\]
 and
\[
(-1)^m a_{2m+1} (X^1)^{2m+1} dp, \qquad  m \in N-\{0\},
\]
where $ a_{2m+1} $ is related to the so called  Catalan numbers
and   is given by \be \label{207q}
 a_{2m+1} =
\frac{2}{m} \left(
\begin{array}{c}
2m-2 \\
m-1
\end{array}
\right) \frac{1}{16^m} , \qquad m \geq 1.
 \ee
The expression (\ref{208}) allows us to write eigenvalue equations
for  observables.
 Let us start with constructing the explicit form of the eigenvalue equation for momentum $ p. $
The general form of this equation is \be \label{31u} p * W_{\bf p
}(q,p)= {\bf p } \cdot W_{\bf p }(q,p), \ee where $ {\bf p } $
denotes the eigenvalue of $p$ and $ W_{\bf p }(q,p)$ is the Wigner
function associated to the eigenvalue $ {\bf p }. $ We find that
\be\begin{array}{lll} \label{310}
 \sigma^{-1}(p)&=&  p +
X^2 + \frac{1}{2} p X^1 X^1 +\frac{1}{8} X^1 X^1 X^2
\\[0.25cm]
 &&\quad -\frac{1}{128}  X^1 X^1 X^1 X^1 X^2 +
\frac{1}{1024} X^1 X^1 X^1 X^1 X^1 X^1 X^2  - \cdots
 \end{array}\ee
 Every
coefficient $ b_{2m+1}$ appearing in the term $
b_{2m+1}\underbrace{X^1 \cdots X^1 }_{2m }X^2$, for $ m > 0 $, can
be expressed in the form 
\be \label{309}
b_{2m+1}=(-1)^{m+1}a_{2m+1},
 \ee
where $ a_{2m+1} $ is defined by  (\ref{207q}).

Much more complicated is to find the general formula of the series
$\sigma^{-1}(W_{\bf p })$ representing the Wigner function $W_{\bf
p }(q,p)$. After long considerations we conclude that   the
function $W_{\bf p }(q,p)$ depends only on $p $ and that the
eigenvalue equation for $W_{\bf p }(p)$ is the infinite
differential equation 
\be \label{315}
 p \cdot \left( W_{\bf p }(p)
- \frac{1}{8}\hbar^2 \frac{d^2 W_{\bf p }(p)}{dp^2} -
\frac{1}{128}\hbar^4 \frac{d^4 W_{\bf p }(p)}{dp^4}  - \cdots
\right)= {\bf p } \cdot  W_{\bf p }(p).
\ee 
The eigenvalue equation for momentum $ p $ is a differential
equation of  infinite degree. There is no general method for
solving such equations. Therefore,
 we decided to look for the solution of the  eigenvalue equation of $p^2$. As
$ p \cdot p= p * p \; $
 the Wigner function fulfilling the equation (\ref{31u}) satisfies also the
relation \be \label{316} p^2*W_{\bf p }(p)= {\bf p^2 } W_{\bf p
}(p). \ee We can see that the above relation is the modified
Bessel equation 
\be \label{317}
 \frac{1}{4}\hbar^2 p^2 \frac{d^2
W_{\bf p }(p)}{dp^2} +\frac{1}{4}\hbar^2 p \frac{d W_{\bf p
}(p)}{dp} +({\bf p^2 }-p^2)W_{\bf p }(p)=0. 
\ee 
The general
solution  of (\ref{317}) is a linear combination  of the following
form 
\be \label{318} 
W_{\bf p }(p)=A \cdot I_{\frac{2i{\bf p}}
{\hbar}}(\frac{2p}{\hbar}) + B \cdot K_{\frac{2i{\bf p}}
{\hbar}}(\frac{2p}{\hbar}) ,
\ee
 with $I_{\frac{2i{\bf p }}{\hbar}}(\frac{2p}{\hbar}) $ 
  the modified Bessel function
with  complex parameter $\frac{2i{\bf p }}{\hbar}$ and 
$K_{\frac{2i{\bf p }}{\hbar}}(\frac{2p}{\hbar})$   the modified
Bessel function of  second kind with parameter $\frac{2i{\bf p
}}{\hbar}$. This solution is defined for arguments $
\frac{2p}{\hbar}>0 $ (see \cite{Sch}). Note that the deformation
parameter $ \hbar $ appears   in the denominator of the argument.
We must be very careful because this fact may cause the
 nonexistence of the series  $ \sigma^{-1}(W_{\bf
p}(p)). $

The function $ A \cdot I_{\frac{2i{\bf p
}}{\hbar}}(\frac{2p}{\hbar}) $ is complex. Its real part grows up
to infinity for $ x \rightarrow \infty $ and, hence, it is not
normalizable. The imaginary part of $ I_{\frac{2i{\bf p
}}{\hbar}}(\frac{2p}{\hbar})$  is proportional to $
K_{\frac{2i{\bf p }}{\hbar}}(\frac{2p}{\hbar})$. Thus, the only
physically admissible solution of (\ref{317}) is \be \label{319}
W_{\bf p }(p)= B \cdot K_{\frac{2i{\bf p
}}{\hbar}}(\frac{2p}{\hbar}).
 \ee
However, solutions defined on the whole $ R $ are required. It is
impossible to define the solution of the equation (\ref{317}) on
the whole axis. The problem is that the modified Bessel function
of  second kind is not defined for the argument value $p=0$.
Moreover, $\lim_{p\rightarrow 0^+} K_{\frac{2i{\bf p
}}{\hbar}}(\frac{2p}{\hbar}) $ does not exist.

Let us assume that the Wigner function $ W_{\bf p }(p) $ is a
generalized function over the Schwartz space $ {\cal S}(p) $ of
test smooth functions tending to $0$ when $ p \rightarrow \pm
\infty$ faster than the inverse of any polynomial.

We define the Wigner function $W_{\bf p }(p)$ as:
\begin{enumerate}
\item
${\bf p}<0$
\be \label{aa}
 W_{\bf p} (p)=
\left\{
\begin{array}{cc}
 \frac{4 }{\pi \hbar} \cosh{\frac{{\bf p} \pi }{\hbar}}
K_{\frac{2i{\bf p }}{\hbar}}(\frac{-2p}{\hbar}) & \mbox{for}
\;\;\; p<0 \\[0.25cm]
0  & \mbox{for} \;\;\; p \geq 0
\end{array}
\right. \ee

\item
 ${\bf p}>0$
 \be \label{bb}
 W_{\bf p} (p)=
\left\{
\begin{array}{cc}
0  & \mbox{for} \;\;\; p \leq 0\\[0.25cm]
\frac{4 }{\pi \hbar} \cosh{\frac{{\bf p} \pi }{\hbar}}
K_{\frac{2i{\bf p }}{\hbar}}(\frac{2p}{\hbar}) & \mbox{for} \;\;\;
p>0
\end{array}
\right. \ee 
\item 
 ${\bf p}=0$ \be \label{cc}
 W_{\bf 0} (p)=
\left\{
\begin{array}{cc}
\frac{2 }{\pi \hbar} K_{0}(\frac{-2p}{\hbar})
& \mbox{for} \;\;\; p<0 \\[0.25cm]
\frac{2 }{\pi \hbar} K_{0}(\frac{2p}{\hbar}) & \mbox{for} \;\;\;
p>0.
\end{array}
\right. \ee
\end{enumerate}
The example analyzed before is a particular choice of $G$ in
 the $ 2 $-form of curvature
\be \label{ex1} R=\pm G^2 X^1X^1 dq \wedge dp, \ee where $ G$ is
some positive constant. The solutions of the equation (\ref{ex1})
are divided in two classes for $R= G^2 X^1X^1 dq \wedge dp$ and
$R=- G^2 X^1X^1 dq \wedge dp$, respectively. For $G \rightarrow 0$
both of them tend to $\delta(p - {\bf p})$.

 The eigenvalue equation for position $q$ takes the form
\be
 \label{.1}
 q*W_{\bf
q}(q,p)={\bf q}W_{\bf q}(q,p) ,
 \ee
where ${\bf q}$ denotes the eigenvalue of the position $q$.
 It can be separated in
two parts: the real part
 \be
\label{.2} q \cdot W_{\bf q}(q,p)={\bf q}W_{\bf q}(q,p) \ee and
the imaginary part \be \label{.3} \frac{1}{2}\hbar\frac{\partial
W_{\bf q}(q,p)}{\partial p} + \frac{1}{96}\hbar^3\frac{\partial^3
W_{\bf q}(q,p)}{\partial p^3} + \cdots =0 .
 \ee

Let us start with the equation (\ref{.3}). Multiplying it by
$\frac{2}{\hbar},$
 introducing a new variable $z=\frac{\sqrt{2} p}{ \hbar}$ and defining
$ w_{\bf q}(q,z)  =\frac{\partial W_{\bf q}(q,z)}{\partial z},$ we
obtain the formula \be \label{.4}
 w_{\bf q}(q,p) + \frac{1}{24}\frac{\partial^2 w_{\bf q}(q,p)}{\partial z^2} + \cdots =0.
\ee This is, in fact, a homogeneous linear differential equation
of  infinite degree. Its solution neither depends on the parameter
$ {\bf q}$ nor the factor  $G$. Since the $\lim_{G \rightarrow 0^{+}}
w_{\bf q}(q,z)  $ must be $0$   the only   admissible solution
of (\ref{.4})
 is $w_{\bf q}(q,z)=0$ for every ${\bf q}$. We see that the Wigner eigenfunction
$ W_{\bf q}(q,z)$ depends only on $q$. From (\ref{.2}) we
immediately obtain that 
\be \label{.5} 
W_{\bf q}(q,z)=\delta(q-{\bf q}) , \qquad \forall {\bf q}. 
\ee 
Similar
considerations can be done for any arbitrary curvature 2-form  $R$ of the form
(\ref{ex1}). The final result will be the same, i.e. the
eigenfunctions $ W_{\bf q}(q)$ depend only on $q $ and have  the
same form like in the case of the flat space ${\mathbb R}^{2n}$
with $ \Gamma=0$.

It is well known \cite{en1,en2} that on a $2$-dimensional
symplectic manifold  all the $*$-products are equivalent. Thus,
for every two products $*_1$ and $*_2$ the relation (\ref{eqv})
holds. It means that the $*$-product considered in this example is
equivalent to the Moyal product (\ref{odw3}). That is true but we
have not algorithm which enables us to construct the operator
$\hat{T}$ from the equation (\ref{eqv}).We cannot transform Wigner
eigenfunctions and eigenvalues  solutions of the eigenvalue
equation
\[
F(p,q) *_1 W_{F1}(p,q)= F_1 \cdot W_{F1}(p,q)
\]
into eigenvalues and eigenfunctions of
\[
F(p,q) *_2 W_{F2}(p,q)= F_2 \cdot W_{F2}(p,q)
\]
though we are aware of the existence of some relation between
them.


\section*{Acknowledgments}

This work has been partially supported by DGI of the Ministerio de
Ciencia y Tecnolog\'{\i}a of Spain (project BMF2002-02000), the
Programme FEDER of the European Community and the Project of the
Junta de Castilla y Le\'{o}n (Spain VA 085/02). One of us (J. T.)
acknowledges a la Secretar\'{i}a del Ministerio de Educaci\'{o}n y
Cultura of Spain for the grant (SB2000-0129) that supports his
stay in the Universidad de Valladolid during this work started. We
thank Prof. M. Przanowski for his interest in this work and his
valuable remarks.


\end{document}